\documentclass[preprint,3p]{elsarticle}

\usepackage{hyperref}
\usepackage{newtxtext} 

\journal{MSSP}

\biboptions{numbers,sort&compress} 

\usepackage{mathptmx}       

\usepackage{makeidx}         
\usepackage{graphicx}        
\usepackage{natbib}
\usepackage{float}
\usepackage{transparent,color}

\usepackage{todonotes}

\usepackage{amsmath,amsfonts,amssymb}

\usepackage{import}

\usepackage{subcaption,diagbox}
\usepackage{xspace}

\usepackage{subdepth} 

\usepackage{capt-of}
\usepackage{bm}

\usepackage{booktabs,multirow,makecell}

\usepackage{pgfplots}
\usepgfplotslibrary{external}
\usetikzlibrary{plotmarks,arrows,shapes.geometric}
\usetikzlibrary{external}
\pgfplotsset{compat=newest}
\usetikzlibrary{calc}

\newlength\fwidth

\definecolor{orange}{rgb}{0.89844,0.37891,0.00391}%
\definecolor{purple}{rgb}{0.36719,0.23438,0.59766}%
\definecolor{gray}{rgb}{0.7,0.7,0.7}%
\definecolor{yellow}{rgb}{0.9922,0.8867,0.5664}%

\definecolor{lightorange}{rgb}{0.89844,0.68359,0.37891}%
\definecolor{lightpurple}{rgb}{0.78516,0.57031,0.97656}%

\definecolor{myblue}{rgb}{0.1211,0.4688,0.7031}%
\definecolor{mygreen}{rgb}{0.1992,0.6250,0.1719}%
\definecolor{darkgreen}{rgb}{0.1445,0.4492,0.1211}%

\definecolor{lightblue}{rgb}{0.4961,0.7070,0.8164}%
\definecolor{lightgreen}{rgb}{0.4805,0.7266,0.5352}%

\newcommand{\ie}{i.e.\,}
\newcommand{\eg}{e.g.\,}
\newcommand{\cf}{cf.\,}
\newcommand{\etal}{et\,al.\,}
\newcommand{\sref}[1]{Section \ref{sec:#1}}
\newcommand{\aref}[1]{\ref{append:#1}}
\newcommand{\eref}[1]{Eq.\ (\ref{eq:#1})}
\newcommand{\fref}[1]{Fig.\ \ref{fig:#1}}
\newcommand{\tref}[1]{Tab.\ \ref{tab:#1}}
\newcommand{\EPMC}{EPMC\xspace}

\newcommand{\DOF}{DOF}

\newcommand{\ee}{\mathrm{e}}
\newcommand{\ii}{\mathrm{i}}

\newcommand{\real}[1]{\operatorname{Re}\left\lbrace #1 \right\rbrace}

\newcommand{\herm}{^{\mathrm H}}
\newcommand{\tra}{^{\mathrm T}}
\newcommand{\pseudoinv}{^{+}}
\newcommand{\e}[2]{\begin{equation} #1 \label {eq:#2} \end{equation}}
\newcommand{\ea}[2]{
	\begin{eqnarray}
	#1 \label {eq:#2} \end{eqnarray}}

\newcommand{\appr}{_\text{appr}}
\newcommand{\exper}{_\text{exp}}
\newcommand{\indexexc}{_k}
\newcommand{\indexmode}{_m}
\newcommand{\indexharm}{_n}

\newcommand{\timevar}{t}
\newcommand{\period}{T}

\newcommand{\displacement}{\mathbf x}
\newcommand{\velocity}{\dot{\mathbf x}}
\newcommand{\acceleration}{\ddot{\mathbf x}}

\newcommand{\ForceVec}{\hat{\mathbf f}}
\newcommand{\forceVec}{\mathbf f}
\newcommand{\forcescalar}{f}

\newcommand{\nlforce}{\mathbf g}

\newcommand{\mass}{\mathbf M}
\newcommand{\damping}{\mathbf D}
\newcommand{\stiffness}{\mathbf K}

\newcommand{\unitvec}{\mathbf{e}}

\newcommand{\scale}{\nu}
\newcommand{\scalerms}{\tilde{\nu}}

\newcommand{\modamp}{a}
\newcommand{\ommod}{\omega}
\newcommand{\Dmod}{\zeta}

\newcommand{\shpmod}{\hat{\mathbf x}}
\newcommand{\shpmodscalar}{\hat{x}}

\newcommand{\shpmodnorm}{\bm\psi}

\newcommand{\omlin}{\omega_0}
\newcommand{\Dlin}{\zeta_0}

\newcommand{\shpmodnormlin}{\bm\phi}
\newcommand{\shpmodnormscalarlin}{\phi}
\newcommand{\shpmodmatrixlin}{\bm\Phi}

\newcommand{\locationexp}{\mathbf C}
\newcommand{\xexp}{\mathbf y}

\newcommand{\numberdof}{N}

\newcommand{\phasemod}{\theta_\mathrm{a}}
\newcommand{\phaselag}{\theta}

\makeindex             

\begin{document}

\begin{frontmatter}
\title{Nonlinear modal testing of damped structures: Velocity feedback vs. phase resonance}

\author[addressILA]{Maren Scheel}

\address[addressILA]{University of Stuttgart, Pfaffenwaldring 6, 70569 Stuttgart, Germany; scheel@ila.uni-stuttgart.de} 

\begin{abstract}
	In recent years, a new method for experimental nonlinear modal analysis has been developed, which is based on the extended periodic motion concept. The method is well suited to experimentally obtain amplitude-dependent modal properties (modal frequency, damping ratio and deflection shape) for strongly nonlinear systems. To isolate a nonlinear mode, the negative viscous damping term of the extended periodic motion concept is approximated by ensuring phase resonance between excitation and response. 
	In this work, an alternative approach to isolate a nonlinear mode is developed and analyzed: velocity feedback. 
	The accuracy of the extracted modal properties and robustness of velocity feedback is first assessed by means of simulated experiments.
	The two approaches phase resonance and velocity feedback are then compared in terms of accuracy and experimental implementation effort.
	To this end, both approaches are applied to an experimental specimen, which is a cantilevered beam influenced by a strong dry friction nonlinearity.	
	In this work, the discussion is limited to single-point excitation. 
	It is shown that a robust implementation of velocity feedback requires the measurement of several response signals, distributed over the structure. An advantage of velocity feedback is that no controller is needed. 
	The accuracy of the modal properties can, however, suffer from imperfections of the excitation mechanism such as a phase lag due to exciter-structure interactions or gyroscopic forces due to single-point excitation.
\end{abstract}


\end{frontmatter}

\section{Introduction}

Modal analysis is a widespread and powerful tool for vibration analysis tasks. It is commonly used for model updating and validation, quality assessment and structural health monitoring, or controller design, just to name a few. Modal analysis methods are well established for linear vibrations, but only very few methods are suited for nonlinear, damped systems \cite{Ewins2000}.
To predict the vibration behavior, such as the decay of free vibrations, the response at resonance or the susceptibility to self-excitation, the quantification of damping is crucial.
Damping is, however, almost always nonlinear. Dry friction and hysteretic material behavior are two important technical examples for nonlinear dissipation.
To facilitate accurate vibration analysis of nonlinear, damped systems, suitable modal analysis methods have been, and still are, subject of active research.

One commonly used concept of a nonlinear mode defines it as periodic motion of a conservative system \cite{Rosenberg1960}.
This notion was extended to damped systems in the extended periodic motion concept (\EPMC) \cite{Krack2015}. In this concept, periodicity is enforced by a mass-proportional negative damping term, which compensates the natural dissipation of the system over one cycle of vibration.
Let us consider the dynamics of the system
\e{
	\mass \acceleration (\timevar) + \nlforce (\displacement (\timevar),\velocity (\timevar)) = \mathbf{0},
}{eqm}
with generalized coordinates $\displacement \in \mathbb{R}^{\numberdof \times 1}$ and the positive definite, symmetric mass matrix $\mass$. The force vector $\nlforce$ contains linear and nonlinear restoring and damping forces. The over-dot indicates the derivative with respect to time $\timevar$. Time-dependence is subsequently dropped for brevity.
The nonlinear mode according to the \EPMC is defined as the family of periodic solutions $\displacement(\timevar) = \displacement (\timevar + \period)$ of the surrogate system
\e{
	\mass \acceleration + \nlforce (\displacement,\velocity) - 2\ommod \Dmod \mass \velocity =  \mathbf{0}.
}{eq_mode}
The modal properties frequency $\ommod = 2\pi / \period$, damping ratio $\Dmod$, and deflection shape (to be defined later) depend on the vibration level. 
This definition of a nonlinear mode is consistent with the linear case under modal damping and continuously extends a corresponding linear mode from the considered equilibrium point. The artificial, negative damping term may distort the modal coupling if more than one linear mode contributes strongly and, at the same time, damping is not light. If the system is forced or self-excited near an isolated primary resonance, the system's dynamics are dominated by a single nonlinear mode (single-nonlinear-mode theory \cite{Szemplinska-Stupnicka1979}). Then, the dynamics are well replicated by a single nonlinear modal oscillator with the modal properties according to the \EPMC \cite{Krack2014a}. 

Recently, another concept for damped systems was introduced, where the nonlinear mode is defined as enforced periodic motion: the phase resonance nonlinear mode \cite{Volvert2021}.
With this concept, not just fundamental resonances but also superharmonic and subharmonic resonances under harmonic forcing are characterized by the phase resonance nonlinear mode. This enhances the predictive capability of a nonlinear mode to those types of resonances.
Periodicity of the motion is enforced by feeding back a filtered, delayed velocity signal. The signal is applied to one degree of freedom (\DOF ), \ie where the external forcing is located. The relevance of the phase resonance nonlinear mode for different forcing scenarios, such as a change of excitation location or self-excitation, is yet unclear. 
Further, phase resonance nonlinear modes have been studied so far only for single and two \DOF -systems with linear damping.

Besides rigorous theoretical concepts, experimental data is needed to validate and update model \cite{Kerschen2006,Noel2017b}. Numerous approaches for nonlinear modal testing have been developed in recent years. Nonlinear modal testing refers to obtaining amplitude-dependent modal properties from experimental data.
Several researches have proposed to extract modal properties from free decay measurements, either using hammer excitation \cite{Kuether2016,Sracic2012,Stephan2017} or force appropriation with subsequent removal of the excitation \cite{Peeters2011a,Heller2009,Londono2015}. The signal length of a free decay is inherently limited and depends on the damping. This limits the resolution of the identified modal properties and increases the sensitivity to noise. 

Standard linear modal testing algorithms can be utilized for nonlinear modal testing if force-controlled frequency response functions (FRFs) at multiple levels are obtained (\eg \cite{Lin1993}). Force control, however, fails for multivalued FRFs. Response-controlled FRFs have been suggested as remedy \cite{Link2011,Karaagacli2021}, as response control yields linearized FRFs and is readily available in commercial modal analysis software. However, testing at many forcing or response levels is a time-consuming procedure. It raises the risk of damage during testing and can cause challenging time-variance due to thermal effects.

With phase resonance testing, the amplitude-dependent modal properties are directly obtained, without the need of measuring FRFs, which reduces the required measurement duration. 
Peeters \etal \cite{Peeters2011a} were the first to extend the concept of phase resonance to nonlinear systems. In this seminal work, phase resonance was achieved by manually tuning the excitation frequency. The subsequent free decay was analyzed to extract the modal properties. To avoid the aforementioned drawbacks of free decay measurements and cumbersome manual tuning, phase resonance testing with steady-state excitation was suggested using a phase-locked loop controller (PLL) \cite{Denis2018,Peter2017} or control-based continuation \cite{Renson2016b,Renson2017}. These studies show that for most structures, it is sufficient to control the phase associated with the fundamental harmonic response and force. It was further shown that phase resonance testing using a PLL controller is well suited to experimentally implement the \EPMC \cite{Scheel2018}. For example, the modal properties of a cantilevered beam with strong friction nonlinearity, called RubBeR, were extracted with good accuracy \cite{Scheel2020b}. 

The PLL controller is a control scheme, which is not implemented in standard vibration testing packages. Its use requires tuning of the control parameters, which increases the effort of setting up the experiment. Bad tuning might lead to very slow phase locking, such that the experiment duration is practically infeasible, or even lead to unstable control loops. Moreover, the control parameters must be tuned for each specimen.

In this work, an alternative approach to experimentally implement the \EPMC is suggested that does not require a controller. To this end, velocity feedback is employed that aims at mimicking the effect of the negative mass-proportional damping term of the \EPMC. To minimize the need of experimental equipment, the discussion is limited to single-point forcing. Two versions for velocity feedback are proposed (\sref{enma}): either only one velocity signal or a weighted sum of multiple velocity signals is fed back.
The accuracy of the two implementations are determined by means of simulated experiments and comparing with the modal properties according to the \EPMC (\sref{virtual}).
Then, one promising velocity feedback approach is compared with phase resonance testing using PLL by applying both methods to the RubBeR test rig (\sref{rubber}). 


\section{Nonlinear modal testing based on the \EPMC}\label{sec:enma}

To implement the \EPMC experimentally, the negative damping term in \eref{eq_mode} is interpreted as excitation force (on the right-hand side of the equation), $\forceVec = 2\ommod \Dmod \mass \velocity$. In an experiment, it is impossible to apply such a distributed force to a structure. Instead, forces at only a finite number of points are applied. In fact, excitation at only one location is commonly preferred due to the limited availability of vibration exciters in a laboratory. Further, utilizing only one exciter reduces potential effects on the structure's impedance due to the attached exciter. Therefore, the discussion is limited to force appropriation applied at only one location, $\forceVec\appr  = \unitvec\indexexc \forcescalar $. $\unitvec\indexexc \in \mathbb{R}^{\numberdof \times 1}$ is the vector locating the excitation \DOF\ $k$. In the following, it is assumed that one generalized coordinate coincides with the direction of the force excitation. Then, $\unitvec\indexexc$ is a Boolean vector with only one non-zero entry.

In this section, two approaches of force appropriation based on the \EPMC are presented. First, phase resonance testing is briefly summarized, which was originally derived in \cite{Scheel2018} (\sref{phase_resonance}). Then, velocity feedback is proposed in \sref{velocity_feedback}.

Once the mode is excited in the experiment, using one of the two approaches, the resulting periodic motion and excitation force are recorded. The modal properties are then identified from the recorded signals \cite{Scheel2018}. The modal frequency $\ommod$ and $N_\mathrm{h}$ Fourier coefficients of the deflection shape $\shpmod_n \in \mathbb{C}^{\numberdof \times 1}$ are extracted directly from the measured time signals with $\displacement(\timevar) = \real { \sum\limits_{n=0}^{N_\mathrm{h}} \shpmod\indexharm \ee^{\ii n \ommod \timevar}}$. $\ii$ is the imaginary unit.

To extract the modal damping ratio $\Dmod$, the excitation power is balanced with the dissipated power. If only the fundamental harmonic component of the excitation is controlled, \ie for a purely harmonic excitation signal, higher harmonics in the excitation force are caused by the shaker-structure interactions. Then, the higher harmonics are not useful to characterize the structure under test. Therefore, it is proposed to use the (controlled) fundamental harmonic component only for the quantification of the damping.
The active excitation power $P_1$ is associated with the first Fourier coefficient of the measured excitation force, $\hat{\forcescalar}_1 \in \mathbb{C}$, and the first Fourier coefficient of the response at the exciter location, $\shpmodscalar\indexexc{}_{,1} = \unitvec\indexexc\tra \shpmod_1$. The response at the exciter location, $x\indexexc = \unitvec\indexexc\tra \displacement$, is subsequently referred to as drive-point response.
The modal damping ratio is then \cite{Scheel2018}
\e{
	\Dmod = \dfrac{P_1}{\ommod^3 \modamp^2}  = \dfrac{\real{\ii \ommod \shpmodscalar\indexexc{}_{,1} \bar{\hat{\forcescalar}}_1  }}{2 \ommod^3 \modamp^2}.
}{dmod}
$\bar{(\bullet)}$ indicates the complex conjugate. $\modamp > 0$ is the magnitude of the modal amplitude, \ie the scaling factor between the first Fourier coefficient of the mass-normalized deflection shape, $\shpmodnorm_1$, and the unscaled shape, $\shpmod_1 = \modamp \shpmodnorm_1$. For this mass-normalization step, the mass matrix is estimated using linear mass-normalized mode shapes,
\e{
	\mass\exper = (\shpmodmatrixlin\exper\tra)\pseudoinv \shpmodmatrixlin\exper\pseudoinv.
}{mass}
${(\bullet)}\tra$ is the transpose, and ${(\bullet)}\pseudoinv$ is the pseudo-inverse of a matrix. 
The columns of $\shpmodmatrixlin\exper \in \mathbb{R}^{\numberdof\exper \times \numberdof_\mathrm{m}}$ are $\numberdof_\mathrm{m}$ linear mode shapes. These are obtained with standard linear experimental modal analysis, conducted at low vibration level, at which the vibration behavior is (nearly) linear. To this end, the response is measured at $\numberdof\exper \geq \numberdof_\mathrm{m}$ \DOF s.
With this, the magnitude of the modal amplitude is obtained with
\e{
	\modamp^2 = \shpmod_1\herm \mass\exper \shpmod_1.
}{modamp}
${(\bullet)}\herm$ is the Hermitian transpose.

\subsection{Force appropriation using phase resonance}\label{sec:phase_resonance}

To derive force appropriation using a phase resonance criterion, the excitation force is rewritten according to the EPMC at the $k$-th \DOF\ and for the $n$-th harmonic \cite{Scheel2018},
\e{
	\hat{\forcescalar}\indexexc{}_,{}\indexharm = 2 \ii n\ommod^2 \Dmod \left( M\indexexc{}\indexexc \shpmodscalar\indexexc{}_,{}\indexharm + \sum\limits_{j\neq k}^{} M\indexexc{}_j \shpmodscalar_{j,}{}\indexharm \right).
}{}
$M\indexexc{}\indexexc$ is a diagonal entry of the mass matrix $\mass$, $M\indexexc{}_j$ with $k \neq j$ an off-diagonal entry.
Assuming that the mass matrix is diagonal-dominant (which is a valid assumption for slender structures),  the force can be approximated with \cite{Scheel2018}
\e{
	\hat{\forcescalar}\indexexc{}_,{}\indexharm \approx 2 \ii n\ommod^2 \Dmod M\indexexc{}\indexexc \shpmodscalar\indexexc{}_,{}\indexharm \propto \ii \shpmodscalar\indexexc{}_,{}\indexharm \quad \forall \, n, k.
}{}
This is the \textit{local} phase resonance criterion, \ie a $\pi$/2 phase lag between drive-point displacement response and excitation force.

Previous studies show that ensuring phase resonance for the fundamental harmonic component only is sufficient for many structures \cite{Peeters2011a,Denis2018,Peter2017,Scheel2018,Renson2016b,Renson2017,Scheel2020b}.
Controlling higher harmonics in the excitation force increases the control effort substantially. Moreover, higher harmonics are not expected to have a significant effect as long as they do not engage into resonance with another mode.
Applying the force at only one location, the resulting force is
\e{
	\forcescalar = \underbrace{\real{\hat{\forcescalar}_1\ee^{\ii\ommod \timevar}}}_{\text{controlled}} + \underbrace{\real{\sum\limits_{n=2}^{\infty} \hat{\forcescalar}\indexharm\ee^{\ii n\ommod \timevar}}}_{\text{uncontrolled}}.
}{}
This force comprises the controlled harmonic and uncontrolled higher harmonics, caused by the interaction between exciter and structure.
The amplitude of the fundamental harmonic is chosen by the user to achieve varying vibration levels. In practice, this can simply be achieved by varying the input gain of the exciter and let the actual forcing and vibration level be a result of the system dynamics.

The phase lag can be conveniently ensured using a PLL controller \cite{Denis2018,Peter2017,Scheel2018}. The implementation used in this work (see \fref{scheme_exper_flow_pll}) is described in more details in  \cite{Scheel2020b}, and a sketch of the controller is given in \aref{pll}.
 
\begin{figure}
	\centering
	\begin{subfigure}{0.4\textwidth}
		\centering
		\def\svgwidth{\textwidth}
		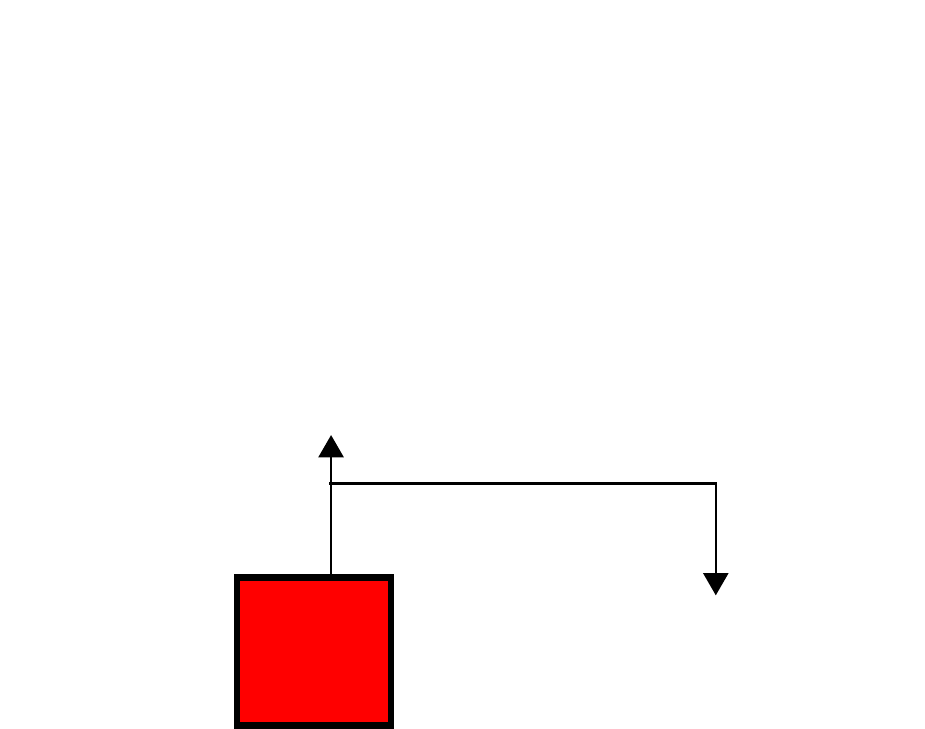
		\caption{}\label{fig:scheme_exper_flow_pll}
	\end{subfigure}
	\begin{subfigure}{0.5\textwidth}
		\centering
		\def\svgwidth{1.9\textwidth}
		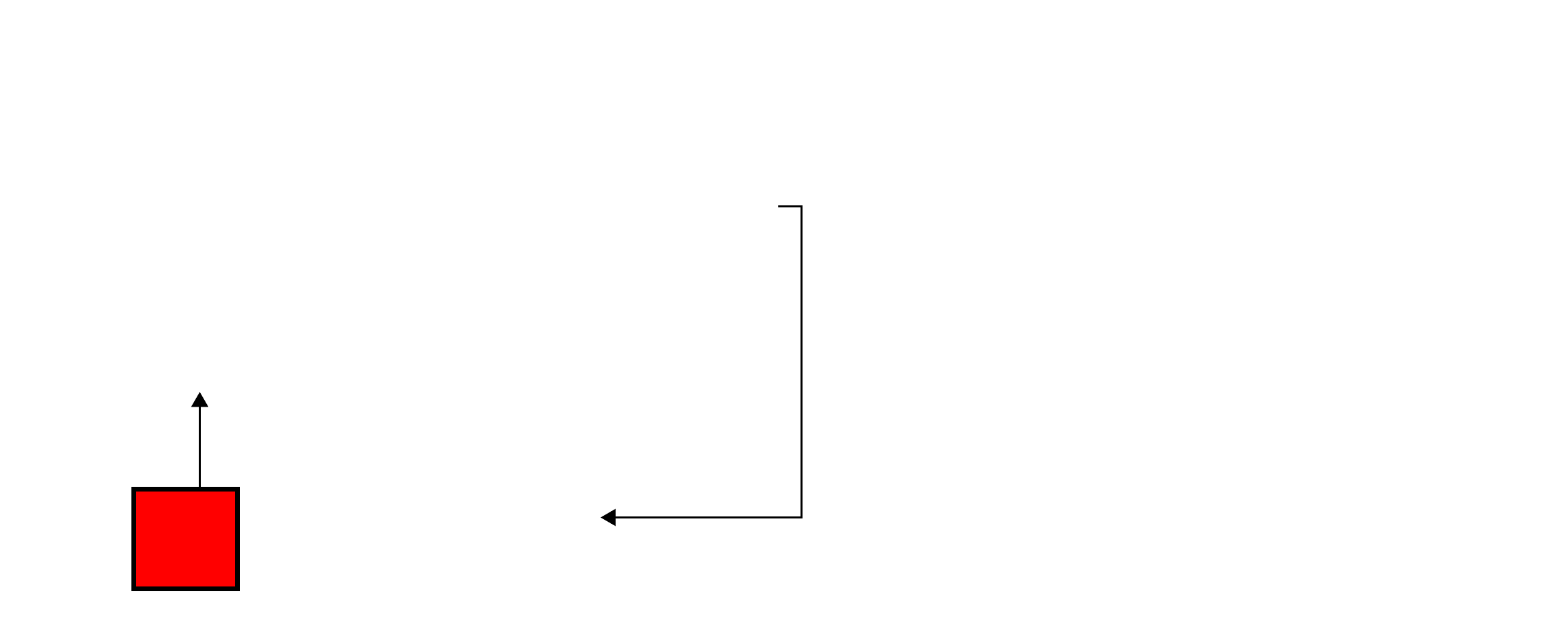
		\caption{}\label{fig:scheme_exper_flow_vfb}
	\end{subfigure}
	\caption{Two approaches for force appropriation based on the \EPMC: (a) phase resonance, (b) velocity feedback with the two weighting schemes single-velocity feedback (sfb) and multi-velocity feedback (mfb). } \label{fig:scheme_exper_flow}
\end{figure}

\subsection{Force appropriation using velocity feedback}\label{sec:velocity_feedback}

The negative damping term of the \EPMC is a particular form of velocity feedback.
In the following, two different schemes are derived for how to approximate this term in an experiment. To this end, two particular challenges inherent with velocity feedback are addressed: the risk of instability and ensuring that the mode of interest is isolated (rather than letting the system decide).

With the weighting matrix $\mathbf{W}$, which is developed in the following, the resulting equation of motion is
\e{
	\mass \acceleration + \nlforce(\displacement, \velocity) - \scale \mathbf{W} \velocity =  \mathbf{0}.
}{eqm_vfb}
$\scale>0$ is a scaling factor that determines the forcing level.

In order to excite a motion with velocity feedback, the velocity must be non-zero. To initialize motion of the system, it is first excited with a frequency that is close to the mode of interest. The linear modal frequency $\omlin$ of the mode of interest is readily obtained with linear modal testing at low levels. Thus, the initialization frequency $\omega_\text{init} = \omlin$ is proposed. After transients have decayed, the excitation is switched to velocity feedback (see \fref{scheme_exper_flow_vfb}).

\subsubsection{Single velocity feedback}

Since only a limited number of sensors is generally available, the motion of only a limited number of \DOF\ is known. 
The simplest implementation for velocity feedback is to only feed back the drive-point velocity, \ie
\e{
	\mathbf{W}_{\text{sfb}}  = \unitvec\indexexc \unitvec\indexexc\tra.
}{s_vel_force}
This excitation scheme, hereafter referred to as \textit{single-velocity feedback} (sfb), causes a negative damping force at one \DOF .

The nonlinear method is required to be consistent with modal testing of linear systems. In the following, single-velocity feedback is analyzed in terms of the modes of the linearized system for small vibrations around the considered equilibrium point. With the mode shape matrix $\shpmodmatrixlin$ of the linearized system, the single-velocity feedback term is $-\scale \shpmodmatrixlin\tra \mathbf{W}_\text{sfb} \shpmodmatrixlin$. The diagonal entries of this fully populated matrix are $-\scale \shpmodnormscalarlin_{i,k}^2 \, \, \text{for} \, \, i =1,...,\numberdof$. $\shpmodnormscalarlin_{i,k}$ is the entry of $\shpmodmatrixlin$ that is associated with the $i$-th mode and the $k$-th \DOF . $\scale \shpmodnormscalarlin_{i,k}^2$ is either positive or zero in case the excitation location coincides with a vibration node of the $i$-th mode.
Assuming light damping, the off-diagonal terms have only secondary effect (see \aref{perturbation}).
Single velocity feedback thus leads to effectively negative damping, applied to all linear modes. Moreover, the modes are coupled. In comparison, the negative damping term of the \EPMC is mass-proportional. It therefore also leads to negative damping of all modes, but does \emph{not} couple the linear modes in the case of modal damping.

Due to mode coupling and negative damping of (nearly) all modes, all modes potentially respond when switching to single-velocity feedback. It is therefore crucial to determine which mode responds.
To this end, a stability analysis is performed with perturbation calculus for the linearized system with modal damping subjected to velocity feedback (see \aref{perturbation} for more details). 
From the stability analysis it follows that the strength of the negative damping (caused by the velocity feedback) depends on the forcing level, \ie the scaling factor $\scale$. The one mode responds, whose cumulative damping (positive modal damping of the mode and negative damping of velocity feedback) switches to negative damping (being excited) for the lowest excitation level.
With single-velocity feedback, this happens first for the mode where $\scale > 2 \Dlin{}_,{}_i \omlin{}_,{}_i / \shpmodnormscalarlin_{i,k}^2$ holds. This ratio is hereafter referred to as critical self-excitation level. According to dynamical systems theory, this condition corresponds to a Hopf bifurcation (assuming sub-critical modal damping ratio).
The condition depends on the modal properties of the linearized system, frequency $\omlin{}_,{}_i$ and damping ratio $\Dlin{}_,{}_i$, as well as the exciter location through the term $\shpmodnormscalarlin_{i,k}$. 
Velocity feedback is useful to isolate the mode whose Hopf bifurcation is reached first (for the lowest $\scale$).
If another mode than the mode of interest meets the critical self-excitation level for a lower excitation level, this mode responds as soon as velocity feedback is switched on.
As soon as another Hopf bifurcation is reached, more than one mode are excited. Then, a dependence on the initial conditions and nonlinear modal interactions can be expected.

To visualize the critical self-excitation level, the example of a cantilevered Euler-Bernoulli beam is considered. Equal damping is assumed for all modes. Then, the critical self-excitation level depends only on the beam's boundary conditions. Computing the beam's modal properties, it can be shown that the first mode's properties yield the smallest critical self-excitation level for excitation in the right half of the beam (the free end). For this range (indicated with dark blue in \fref{beam_linear}), the first mode responds if the system is excited in this range and single-velocity feedback is switched on (irrespective of the initialization frequency).
If the exciter is located in the light blue or turquoise area of \fref{beam_linear}, the second or third mode responds, respectively. For exciter locations close to the clamping (white area), higher modes respond.

This behavior of single-velocity feedback for linear systems can readily be predicted using the mobility drive-point FRF. The reciprocal of the critical self-excitation level is equal to the absolute value of the FRF at resonance of the corresponding mode. Therefore, given the exciter location, the mode with the highest peak in the drive-point FRF responds when using single-velocity feedback.

\begin{figure}
	\centering
	\def\svgwidth{0.5\textwidth}
\begingroup%
  \makeatletter%
  \providecommand\color[2][]{%
    \errmessage{(Inkscape) Color is used for the text in Inkscape, but the package 'color.sty' is not loaded}%
    \renewcommand\color[2][]{}%
  }%
  \providecommand\transparent[1]{%
    \errmessage{(Inkscape) Transparency is used (non-zero) for the text in Inkscape, but the package 'transparent.sty' is not loaded}%
    \renewcommand\transparent[1]{}%
  }%
  \providecommand\rotatebox[2]{#2}%
  \newcommand*\fsize{\dimexpr\f@size pt\relax}%
  \newcommand*\lineheight[1]{\fontsize{\fsize}{#1\fsize}\selectfont}%
  \ifx\svgwidth\undefined%
    \setlength{\unitlength}{314.89984708bp}%
    \ifx\svgscale\undefined%
      \relax%
    \else%
      \setlength{\unitlength}{\unitlength * \real{\svgscale}}%
    \fi%
  \else%
    \setlength{\unitlength}{\svgwidth}%
  \fi%
  \global\let\svgwidth\undefined%
  \global\let\svgscale\undefined%
  \makeatother%
  \begin{picture}(1,0.18096383)%
    \lineheight{1}%
    \setlength\tabcolsep{0pt}%
    \put(0,0){\includegraphics[width=\unitlength,page=1]{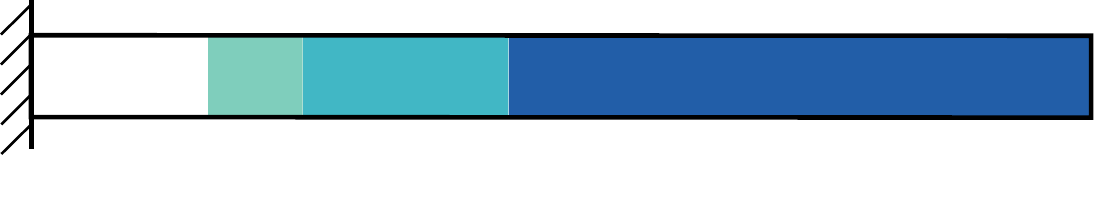}}%
    \put(0.64779158,0.09820812){\color[rgb]{0,0,0}\makebox(0,0)[lt]{\lineheight{1.25}\smash{\begin{tabular}[t]{l}mode 1\end{tabular}}}}%
    \put(0.30284838,0.09820812){\color[rgb]{0,0,0}\makebox(0,0)[lt]{\lineheight{1.25}\smash{\begin{tabular}[t]{l}mode 2\end{tabular}}}}%
    \put(0.27414319,0.0151465){\color[rgb]{0,0,0}\makebox(0,0)[lt]{\lineheight{1.25}\smash{\begin{tabular}[t]{l}mode 3\end{tabular}}}}%
    \put(0,0){\includegraphics[width=\unitlength,page=2]{Fig2.pdf}}%
  \end{picture}%
\endgroup%

	\caption{Indication of which mode responds of a cantilevered Euler-Bernoulli beam to single-velocity feedback, depending on the exciter location.}
	\label{fig:beam_linear}
\end{figure}

\subsubsection{Multi-velocity feedback}

For a given exciter location, single-velocity feedback is expected to isolate only a single mode. One has to change the exciter location for each mode of interest, and it might be practically impossible to excite higher modes. To isolate multiple modes with the same exciter location, another weighting scheme for velocity feedback is proposed: \textit{multi-velocity feedback} (mfb). Here, an estimate of the modal velocity of the mode of interest is fed back,
\e{
	\mathbf{W}_{\text{mfb}} = \unitvec\indexexc \shpmodnormlin\tra\indexmode{}_,{}\exper \mass\exper \locationexp = \unitvec\indexexc \unitvec\indexmode\tra \shpmodmatrixlin\exper\pseudoinv \locationexp.
}{w_vel_force}
$\shpmodnormlin\indexmode{}_,{}\exper$ is the $m$-th column of $\shpmodmatrixlin\exper$, with $m$ being the index of the mode of interest. The matrix $\locationexp$ locates the subset of measured \DOF s $\xexp$, which are used for the velocity feedback, $\xexp \in \mathbb{R}^{N_{\text{exp}} \times 1}$ among all generalized coordinates $\displacement$. Thus, $\xexp = \locationexp \displacement$.
$\unitvec\indexmode \in \mathbb{R}^{\numberdof_\mathrm{m} \times 1}$ is a Boolean vector with the only non-zero entry at the $m$-th entry.
Since $\mathbf{W}_{\text{mfb}}$ is non-symmetric, multi-velocity feedback introduces gyroscopic forces, additionally to damping forces. 

Analyzed in the modal coordinates of the linearized system, multi-velocity feedback leads to the velocity-proportional term $-\scale \shpmodmatrixlin\tra \mathbf{W}_\text{mfb} \shpmodmatrixlin$.
This matrix is visualized in \fref{modal_force_weighted} for the case that the first $\numberdof_\mathrm{m}$ modes are used for the estimation of the mass matrix. The column associated with the $m$-th mode (which is part of the identified $\numberdof_\mathrm{m}$ modes) is fully populated, \ie $\shpmodmatrixlin\tra \unitvec\indexexc$. The entries of the columns associated with the other identified modes are zero. Columns associated with higher, not identified modes are fully populated.
Assuming light damping, the effect of the diagonal entries of $\shpmodmatrixlin\tra \mathbf{W}_\text{mfb} \shpmodmatrixlin$ is dominating (see \aref{perturbation}). The ($m,m$) diagonal entry must be positive to ensure that the mode of interest is indeed excited (and not additionally damped). Therefore, $\shpmodnormlin\indexmode{}_,{}\exper$ has to multiplied with either -1 or 1 to ensure that its $k$-th entry $\shpmodnormscalarlin_{m,k}$ is positive.

\begin{figure}
	\centering
	\def\svgwidth{0.35\textwidth}
	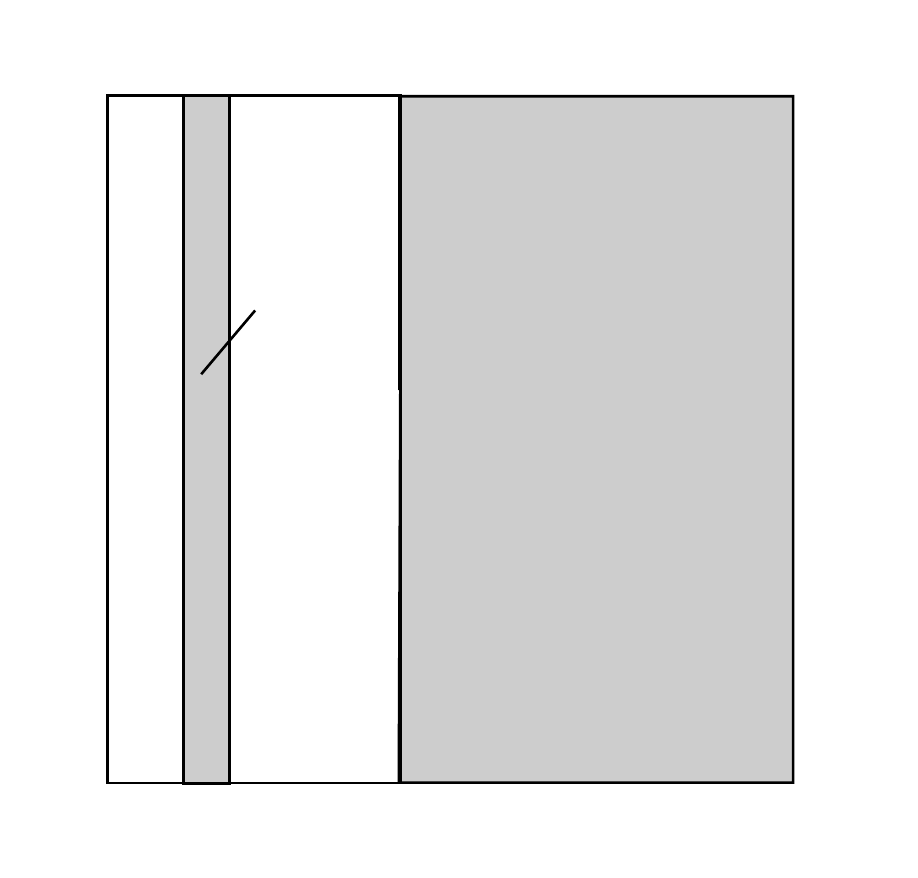
	\caption{Structure of the matrix $\shpmodmatrixlin\tra \mathbf{W}_\text{mfb} \shpmodmatrixlin$.}
	\label{fig:modal_force_weighted}
\end{figure}

With this simplified, single-point implementation of multi-velocity feedback, all modes exhibit additional non-modal damping and gyroscopic forces. Thus, the modes are coupled. The other identified modes (with zero-columns in \fref{modal_force_weighted}) are coupled with higher modes and with the mode of interest and can therefore be indirectly excited.
Extending velocity feedback to multi-point excitation, it would be possible to decouple the other identified modes from the mode of interest and to eliminate gyroscopic forces. Coupling with higher, not identified modes would remain. Multi-point excitation, however, increases the demand for experimental equipment considerably and is out of the scope of this work.

Next, the critical self-excitation level in case of multi-velocity feedback is determined. Repeating the stability analysis (see \aref{perturbation}), this level is $\scale > 2 \Dlin{}_,{}_m \omlin{}_,{}_m / \shpmodnormscalarlin_{m,k}$ for the mode of interest. For higher, not identified modes, the critical self-excitation level is $\scale > 2 \Dlin{}_,{}_i \omlin{}_,{}_i / (\shpmodnormlin_i\tra \mathbf{W}_\text{mfb} \shpmodnormlin_i)$. The other identified modes do not exhibit \emph{modal} damping through velocity feedback. 
Generally, the Hopf bifurcation occurs at lower excitation levels for low-frequency modes that are lightly damped. If the modes not included in the mass-matrix estimation are of higher frequency and of similar (or higher) damping, the critical self-excitation level is reached first for the mode of interest. 

To identify the $m$-th nonlinear mode, two requirements must be met: First, a sufficient number of $\numberdof_\mathrm{m} > m$ linear modes must be identified to ensure that the critical self-excitation level of the mode of interest is reached for sufficiently low excitation levels. Second, the exciter location must not be a vibration node, \ie $\shpmodnormscalarlin_{m,k} \neq 0$. Then, the $m$-th nonlinear mode can be identified with multi-velocity feedback for all exciter locations, which are not vibration nodes.
For the example of the linear cantilevered beam under uniform modal damping, any mode can now be isolated with excitation at the tip (not just the first mode). Analogously, the first mode can be excited with excitation at the left half of the beam. This will be exemplified later in \sref{virtual_wfb}.

\subsubsection{Tracking amplitude-dependent properties and stabilizing unstable motion}

The objective of nonlinear modal testing is to track the amplitude-dependence of the modal properties. Therefore, it must be possible to excite different vibration levels in a systematic way. Depending on the system, this might not be straightforward when using velocity feedback, for example for a linear single \DOF\ system with viscous damping $d$. To achieve periodic motion for this system, the forcing must cancel the damping term, $\scale \dot{x} = d \dot{x}$, which is only fulfilled for exactly one value $\scale$. Then, a conservative system is obtained, whose vibration level depends on the initial conditions. It is therefore not possible to achieve different levels by simply varying the forcing level $\scale$ in a monotonous manner.

Instead, rescaling the forcing term, \ie $\scalerms \dot{x} / \|\dot{x}\| $, is proposed. Then, the damping force is compensated with $d =\scalerms / \|\dot{x}\|$. For constant damping, the vibration level $\|\dot{x} \|$ is now defined by $\scalerms$, \ie $\|\dot{x} \| = \scalerms / d$.
With this scaling, the amplitude-dependence of the modal properties can be tracked by varying $\scalerms$. Here, $\|\bullet \|$ indicates a general amplitude measure. One measure that is readily implemented experimentally is discussed in \sref{practical}. 

Further, dividing by the amplitude is a deliberate 'non-linearization' of the damping term.
The intention of this term is to achieve asymptotic stability of the periodic motion (at a certain level), instead of neutral stability in the linear case. Further, it is intended to achieve asymptotic stability, instead of instability, also in the nonlinear case with decreasing damping. Specifically, this is relevant for example for friction-damped systems at large vibration levels. In case the vibration amplitude grows for a given value $\scalerms$, the excitation is reduced by dividing by $\|\dot{x} \|$. The empirical results of this work verify the proposition of asymptotic stability.
With the proposed scaling, the otherwise unstable modal motion of a friction-damped system at large vibration level can be measured in an experiment.

\subsubsection{Practical aspects}\label{sec:practical}

To summarize, force appropriation with single-velocity feedback is proposed as applying the single-point force
\e{
	\forcescalar_{\text{sfb}} = \dfrac{\scalerms}{\|\dot{x}\indexexc \|} \dot{x}\indexexc 
}{}
and with multi-velocity feedback as applying
\e{
	\forcescalar_{\text{mfb}} = \dfrac{\scalerms}{\| \shpmodnormlin\tra\indexmode{}_,{}\exper \mass\exper \dot{\xexp} \|} \shpmodnormlin\tra\indexmode{}_,{}\exper \mass\exper \dot{\xexp}.
}{}
The signal flow for both cases is sketched in \fref{scheme_exper_flow_vfb}. A summary of the methods are given in \tref{methods}.

\begin{table}
	\centering
	\begin{tabular}{l|c|c|c}
		\toprule
		Method & Phase resonance testing & sfb & mfb\\
		\midrule[0.8pt] 
		\makecell[l]{Number of response locations \\ used for the excitation signal's definition}   & 1 & 1 & $> 1$\\
		Controller needed & yes & no & no\\
		Initialization needed & no & yes & yes\\
		Determination of which mode is excited & \makecell{center frequency \\ of PLL controller \cite{Scheel2020b}}  & exciter location & initialization frequency \\
		Sensitivity to the excitation system & low & high & high\\
		\bottomrule
	\end{tabular}
	\caption{The three approaches used in this work.}\label{tab:methods}
\end{table}

For the suggested scaling, different amplitude measures can be used. Due to its rather simple experimental implementation, the moving root mean square (rms) amplitude is used in this work.
To estimate this measure in real time, the signal is first multiplied with itself, of which the square root is then taken. To obtain an average value, a first order low-pass filter is used. Care must be taken to choose a suitable time constant.
Generally, it is crucial that the amplitude estimate is fast enough in the experiment. If the vibration amplitude grows quickly but the amplitude estimation lacks, the stabilizing effect of dividing by the amplitude is erased. The time constant of the low-pass filter used in the experiments is chosen such that three times the time constant is equal to the expected period length. 

As vibration sensing equipment, accelerometers are commonly used. Sensors that directly measure the velocity, such as laser Doppler vibrometers, are often of limited availability due to their costs. It is, however, not necessary to directly measure the velocity for velocity feedback. In this work, accelerometers are used to measure the acceleration response in the experiment, and the signal is integrated in real time. 
In the case of single-velocity feedback, the results of using an integrated acceleration signal and using a directly measured velocity signal with laser Doppler vibrometer were compared. The accuracy of the extracted modal properties is similar. 

A common exciter for modal testing is an electromagnetic shaker. This excitation mechanism, however, suffers from a phase lag between driving voltage (shaker input) and resulting excitation force due to the interaction with the structure \cite{McConnellVaroto2008} (see \aref{shaker_model}). The effect of the shaker-caused phase lag is studied in more details using simulated experiments in \sref{virtual_exciter}.

\section{Numerical assessment of accuracy and robustness of velocity feedback}\label{sec:virtual}

In this section, simulated experiments are performed with single-velocity and multi-velocity feedback and the extracted modal properties are compared with the reference properties according to the \EPMC. All simulations are conducted with Matlab/Simulink. The specimen is a cantilevered Euler-Bernoulli beam, discretized with seven elements (see \fref{scheme_linear_virtual_exp} and \tref{beam}). Linear damping is modeled as modal damping and defined as 0.3 \% for the first three modes and 1 \% for higher modes. A friction nonlinearity is introduced with an elastic Coulomb element at the third node from the clamping.

\begin{figure}
	\centering
	\def\svgwidth{0.6\textwidth}
	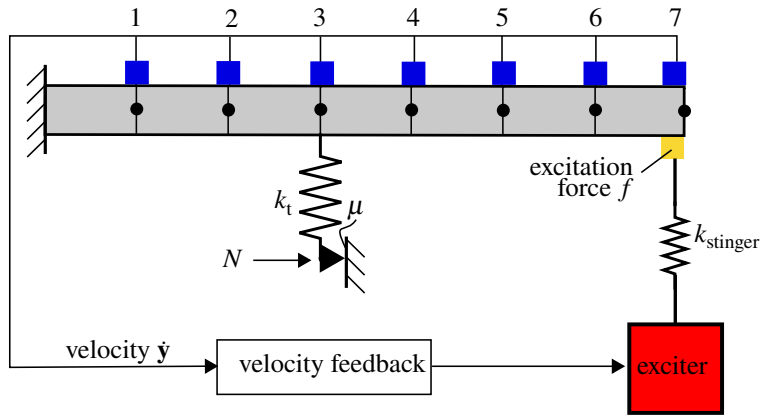
	\caption{Sketch of the simulated experiment with a cantilevered beam with friction nonlinearity.} \label{fig:scheme_linear_virtual_exp}
\end{figure}

\begin{table}
	\centering
	\begin{tabular}{l|l}
		\toprule
		Parameter& Value\\
		\midrule[0.8pt] 
		Young's modulus& $210 \cdot 10^9$ Pa\\
		density & $7830 \text{ kg}/\text{m}^3$\\
		length $l$ & $0.7 \text{ m}$\\
		height & $0.04 \text{ m}$\\
		thickness & $0.06 \text{ m}$\\
		tangential stiffness $k_\mathrm{t}$  & $8 \cdot 10^6$ N/m\\
		slip limit force $\mu N$ & 30 N\\
		\bottomrule
	\end{tabular}
	\caption{Parameters of the specimen for the simulated experiment.}\label{tab:beam}
\end{table}

It is well known that an elastic dry friction nonlinearity leads to a drop in modal frequency with increasing vibration amplitude. The two linear limit cases are \emph{elastic stick} (fixed friction slider) and \emph{full slip} (no deflection in the spring of the elastic friction element). At the same time, the modal damping ratio increases before reaching a maximum and subsequently decreasing again for large vibration amplitudes.
With the chosen parameters, the frequency drop between the limit cases is about 25 \% and 8 \% for the first and second bending mode, respectively. The maximum value of the damping ratio is almost 10 \% for the first mode and  almost 3 \% for the second mode.

The beam's vibration response is recorded as velocity information at all seven nodes in transversal direction. Additionally, the excitation force is recorded. The first three modes are used for the estimation of the mass matrix. These modes lie in a similar frequency range, which was excited in the experiments of \sref{rubber}.

As reference, the modal properties according to the \EPMC are computed using the Matlab code NLvib \cite{KrackGross2019}. The computation is performed with 13 harmonics, which is deemed sufficient as including more harmonics did not influence the modal properties substantially.

First, the accuracy of the extracted modal properties is assessed for single-velocity and multi-velocity feedback in \sref{virtual_sfb} and \sref{virtual_wfb}, respectively. To this end, the experiments are idealized, \ie the excitation signal is directly imposed as force on the structure. 
The effect of the simplifications made for velocity feedback can therefore be assessed: single-point excitation, a limited number of sensors and a limited number of identified modes.
Then, the robustness against the influence of the exciter is studied in \sref{virtual_exciter}. To this end, the exciter's and stinger dynamics are included in the simulation as sketched in \fref{scheme_linear_virtual_exp}. Finally, the robustness against simulated measurement noise is assessed in \sref{virtual_noise}.

\subsection{Accuracy assessment for single-velocity feedback}\label{sec:virtual_sfb}

In \fref{virtual_svel_mode1_exc6}, the modal frequency and the modal damping ratio of the first mode are plotted over the fundamental harmonic amplitude of the beam's tip deflection. The frequency is normalized with the frequency of the linearized system, and the deflection is normalized with the beam's length.
When applying single-velocity feedback with the exciter located at node six (see numbering in \fref{scheme_linear_virtual_exp}), the modal properties are tracked with good accuracy for the full range of vibration levels. Only minor deviations from the reference curves are visible for medium vibration levels.

\begin{figure}
	\centering
	\includegraphics{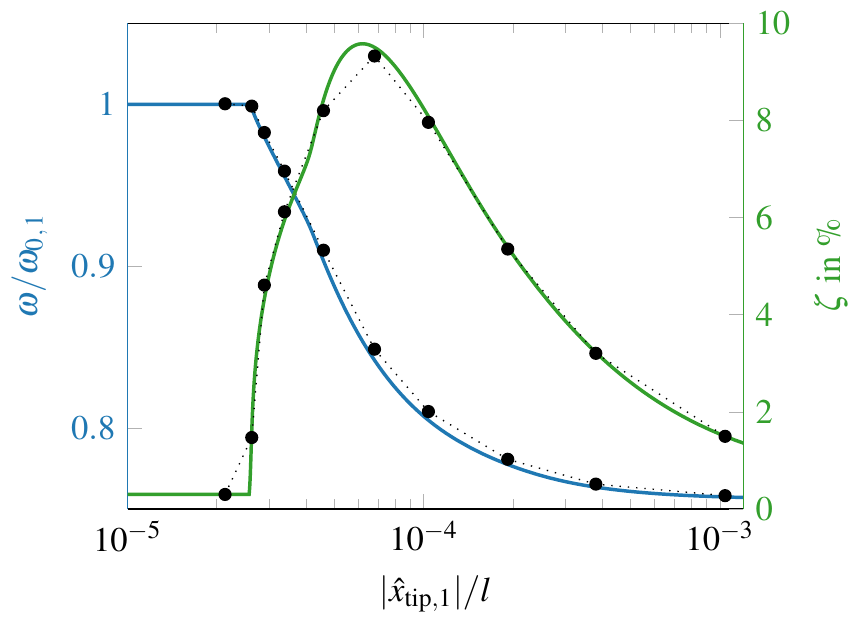}
	\caption{Modal properties of the first mode according to the \EPMC (solid lines) and the result of the simulated experiment with single-velocity feedback for excitation at node six (dots).}
	\label{fig:virtual_svel_mode1_exc6}
\end{figure}

When applying single-velocity feedback at node seven, the first mode (mode of interest) is isolated for very low vibration levels. For a slight increase in excitation level, the responding mode switches to the third mode (see \fref{virtual_svel_mode1_exc7}). 
In the linear regime, the value of the critical self-excitation level $2 \Dlin{}_,{}_3 \omlin{}_,{}_3 / \shpmodnormscalarlin_{3,7}^2$ of the third mode is fifteen times higher than the value for first mode. With increasing vibration level, the critical self-excitation level increases for the first mode due to the strong increase in damping. When the responding mode changes, the modal damping ratio of the first mode is about 4.5 \%, while the frequency and the mode shape have not changed significantly. At the same mechanical energy level, the third mode, however, is still in the linear regime. Thus, the critical self-excitation level for the first mode has increased by about factor 15, while the value for the third mode is constant. Therefore, the third mode is now the responding mode. As can be seen in \fref{virtual_svel_mode1_exc7}, the responding mode switches back to the first mode for very high vibration levels.

If the linear modal damping ratios of the second and third mode are increased from 0.3 \% to 1 \%, the critical self-excitation levels of these modes increase. Now, the first mode is the responding mode for the full range of vibration levels (see \fref{virtual_svel_mode1_exc7_higherdamping}). The extracted modal properties, however, deviate strongly from the reference for medium vibration levels. 
In \fref{virtual_svel_mode1_exc7_fft_a}, the frequency transform of the beam's tip velocity is shown for a vibration level, where the exctracted modal properties are accurate (marked with a in \fref{virtual_svel_mode1_exc7_higherdamping}). Higher harmonics are clearly visible besides the fundamental harmonic frequency. For a vibration level, where the extracted modal properties deviate from the reference (marked with b in \fref{virtual_svel_mode1_exc7_higherdamping}), a strong contribution of a second frequency is visible in the frequency transform (see \fref{virtual_svel_mode1_exc7_fft_b}). This frequency is not an integer multiple of the fundamental frequency. The second frequency is caused by the second mode contributing strongly to the response and leading to a quasi-periodic response.

\begin{figure}
	\begin{subfigure}{0.49\textwidth}
		\includegraphics{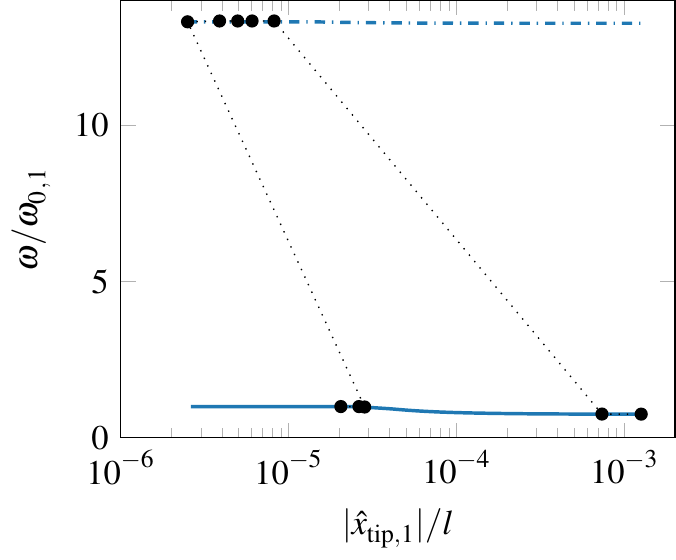}
		\caption{}\label{fig:virtual_svel_mode1_exc7}
	\end{subfigure}
	\begin{subfigure}{0.49\textwidth}
		\centering
		\includegraphics{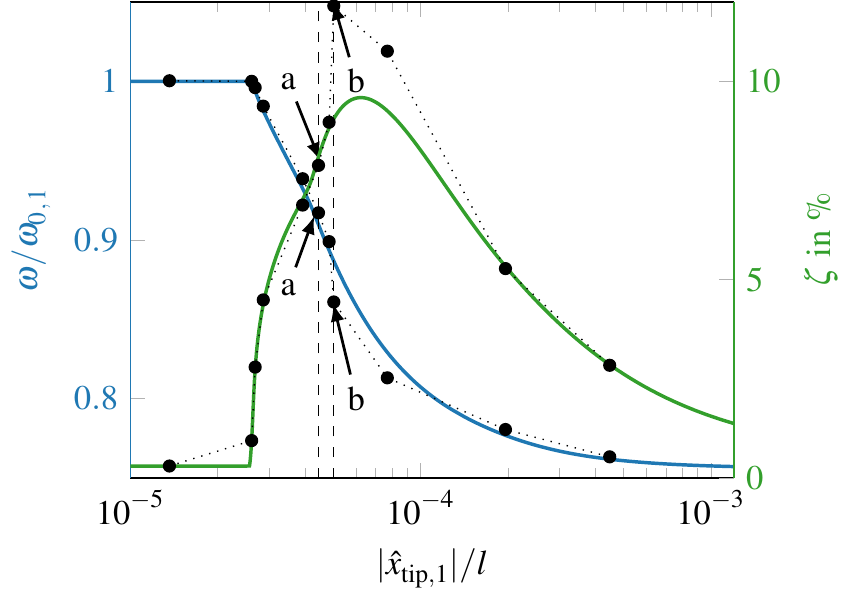}
		\caption{}\label{fig:virtual_svel_mode1_exc7_higherdamping}
	\end{subfigure}
	\caption{Modal properties of the first (solid lines) and third mode (dash-dotted line) according to the \EPMC   and the result of the simulated experiment with single-velocity feedback for excitation at node seven (dots). (a) Linear modal damping ratios of the second and third modes are 0.3 \%. (b) Linear modal damping ratios of the second and third modes are 1 \%.}
\end{figure}

\begin{figure}
	\centering
	\begin{subfigure}{0.45\textwidth}
		\centering
		\includegraphics{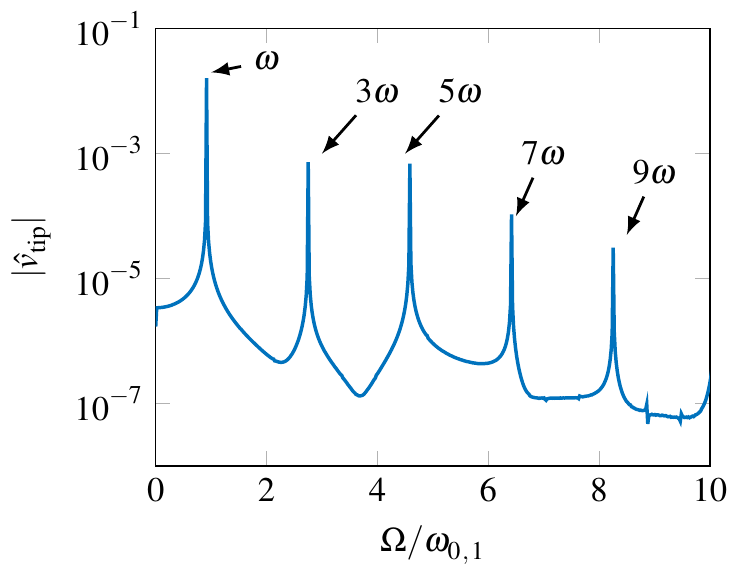}
		\caption{}\label{fig:virtual_svel_mode1_exc7_fft_a}
	\end{subfigure}
	\begin{subfigure}{0.45\textwidth}
		\centering
		\includegraphics{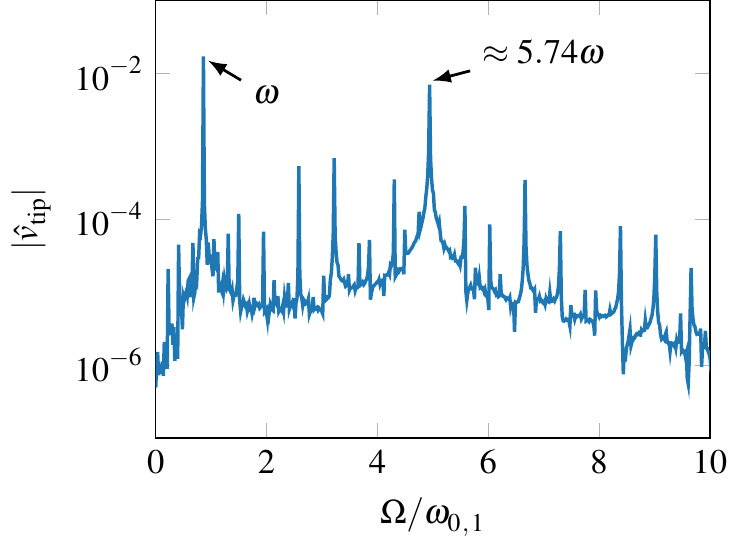}
		\caption{}\label{fig:virtual_svel_mode1_exc7_fft_b}
	\end{subfigure}
	\caption{Frequency transform of the beam's tip velocity for two levels marked in \fref{virtual_svel_mode1_exc7_higherdamping}.}
\end{figure}

Next, single-velocity feedback is applied at node three to isolate the second mode. Here and hereafter, the linear modal damping ratios of the second and third mode are again set to 0.3 \%. The modal properties are tracked for the full range of vibration levels, though deviations are observed for large vibration amplitudes (see \fref{virtual_svel_mode2_exc3}). These deviations are caused by the first mode contributing strongly to the response.

\begin{figure}
	\centering
	\includegraphics{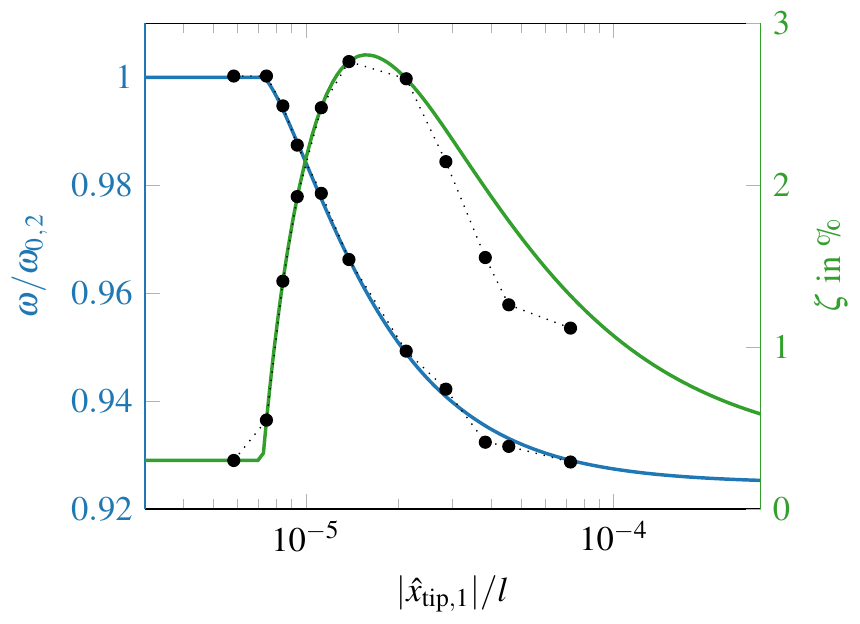}
	\caption{Modal properties of the second mode according to the \EPMC (solid lines) and the result of the simulated experiment with single-velocity feedback for excitation at node three (dots).}
	\label{fig:virtual_svel_mode2_exc3}
\end{figure}

To summarize, single-velocity feedback suffers from a strong sensitivity to the exciter location: The responding mode depends on the exciter location (\cf \tref{methods}). This holds for linear and nonlinear systems. In the latter case, the responding mode can change for varying vibration levels. Therefore, single-velocity feedback is disregarded in the following.

\subsection{Accuracy assessment for multi-velocity feedback}\label{sec:virtual_wfb}

When applying multi-velocity feedback with the exciter located at node seven, the first mode's properties (marked with crosses in \fref{virtual_wvel_mode1}) are tracked with good accuracy for the full range of vibration levels. Exciting at node two, the modal properties are tracked with similar accuracy (marked with dots in \fref{virtual_wvel_mode1}). These results illustrate the increased robustness of multi-velocity feedback towards the choice of exciter location.
The slight deviations from the reference modal frequency could be due to the gyroscopic forces introduced by multi-velocity feedback. Through \eref{dmod}, this also affects the extracted modal damping ratio. Additionally, gyroscopic forces may lead to coupling of (differently damped) modes. 

Applying multi-velocity feedback to isolate the second mode, the modal properties are tracked with high accuracy (see \fref{virtual_wvel_mode2}) for both exciter locations. As the second mode cannot be isolated with single-velocity feedback with excitation at node seven (\cf \fref{beam_linear}), the results underline the robustness of multi-feedback towards exciter locations (\cf \tref{methods}).

\begin{figure}
	\begin{subfigure}{0.49\textwidth}
		\includegraphics{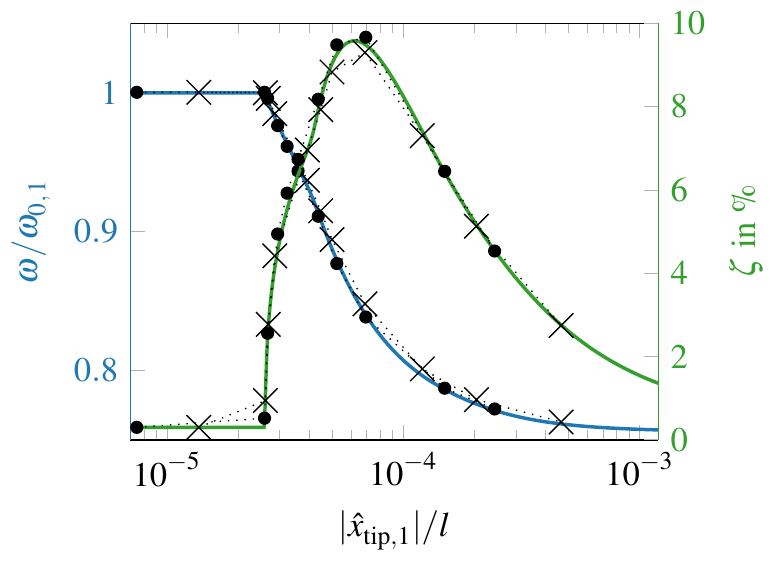}
		\caption{}\label{fig:virtual_wvel_mode1}
	\end{subfigure}
	\begin{subfigure}{0.49\textwidth}
		\centering
		\includegraphics{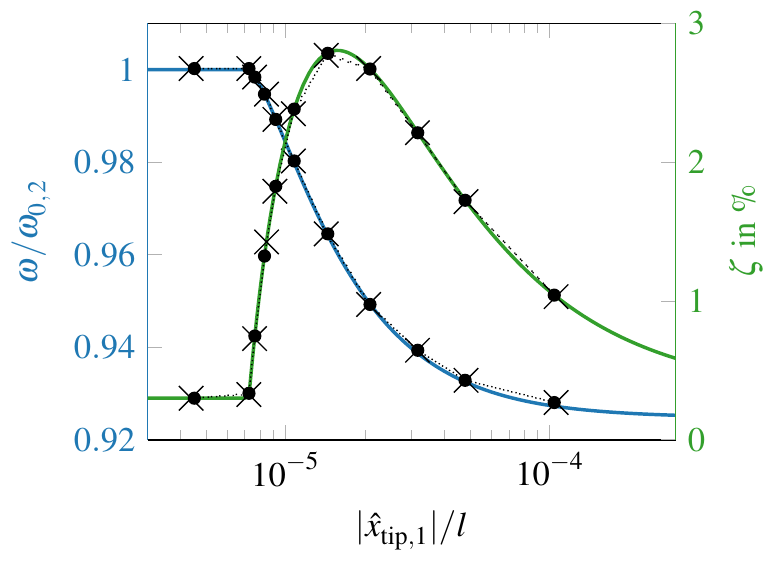}
		\caption{}\label{fig:virtual_wvel_mode2}
	\end{subfigure}
	\caption{Modal properties according to the \EPMC (solid lines) and the results of the simulated experiment with multi-velocity feedback for excitation at node two (dots) and node seven (cross) (a) for the first mode and (b) for the second mode.}
\end{figure}

\subsection{Robustness against the phase lag induced by a conventional exciter}\label{sec:virtual_exciter}

In the following simulations, the excitation mechanism is included. The parameters of the exciter are identified for a Br\"{u}el and Kj\ae r vibration exciter type 4809 and a steel stinger rod modeled as spring (see \aref{shaker_model} for more details). 

Applying multi-velocity feedback at node seven and simulating the exciter's dynamics, the first mode's properties are extracted with similar accuracy as before (\cf \fref{virtual_wvel_mode1_exc7_shaker} and \fref{virtual_wvel_mode1}). 
Only for very low vibration levels, where the system behavior is linear, the extracted modal frequency deviates from the reference. This deviation did not occur in the experiments without exciter model and is caused by a phase lag of the excitation mechanism. 
With the proposed approach for velocity feedback, the \emph{excitation signal} is defined as weighted sum of velocities. A phase lag introduced by the exciter-structure interaction directly affects the mode isolation quality, since the resulting \emph{excitation force} is no longer proportional to the structure's velocity (\cf \tref{methods}).
In comparison, phase resonance testing is not affected by the exciter's phase lag as long as the phase between \emph{measured excitation force} and response is controlled (see \cite{Scheel2018,Balaji2020} for similar simulated experiments using a PLL controller).

As derived in \aref{shaker_model}, the exciter-induced phase lag is most significant for lightly damped systems, if the structure's modal frequency does not coincide with the exciter's armature's modal frequency. For our beam example, damping is low for both low and large vibration levels. The phase lag is, however, only prominent for low vibration levels, not for large levels (see \fref{virtual_wvel_mode1_exc7_shaker_phase}). This is because the full-slip modal frequency is close to the exciter's armature's modal frequency, leading to a nearly real transfer function between excitation force and exciter input. The elastic-stick modal frequency is substantially higher, leading to a complex transfer function and thus a phase lag.
The phase lag could be minimized by exciting close to a vibration node since the phase lag depends on the structure's modal deflection at the exciter location. If the deflection is small (\eg when exciting close to a vibration node), the phase lag is small. Exciting close to a vibration node, however, comes at the cost of a higher forcing level needed to excite similar vibration level, compared to exciting at an anti-node. 

To directly compare the resulting excitation force with the negative damping term of the \EPMC, the forcing terms are transformed to modal coordinates. To this end, the resulting excitation force and the term $2 \Dmod \ommod \mass \velocity$ are projected onto the corresponding mode shape of the linearized system and plotted over the modal deflection (see \fref{virtual_wvel_mode1_exc7_shaker_loops}, for three representative levels). The levels of simulated experiment and \EPMC reference are chosen such that the modal frequencies are equal. 
The general shape of the modal force terms agree well, though deviations in terms of higher harmonic content are visible. It is well known that the exciter-structure interaction cause additional higher harmonics that affect the resulting excitation force \cite{McConnellVaroto2008,Josefsson2006}.

Next, the effect of the exciter on the extracted modal properties is studied for the second mode. Since the maximum damping for this mode is substantially smaller than for the first mode, the phase lag caused by the exciter is more prominent (see \fref{virtual_wvel_mode2_exc7_shaker_phase}). For our beam example, the phase lag first decreases with increasing damping, but then increases again, when damping decreases. For the second mode, both elastic-stick and full-slip modal frequency are higher than the exciter's modal frequency. Further, the modal deflection at the exciter location is similar in both cases. Therefore, the phase lag is mainly affected by the structure's damping.
Analyzing the modal force terms, the effect of the phase lag is visualized by the resulting excitation force deviating strongly from the \EPMC reference (see \fref{virtual_wvel_mode2_exc7_shaker_loops}).

In the extracted modal frequency, the phase lag causes a shift compared to the reference for the full range of vibration levels (see \fref{virtual_wvel_mode2_exc7_shaker}). To further understand the effect of a phase lag on the extracted modal properties, it is quantified for a linear system with modal damping in \aref{phaselag}. The effect is large when the phase lag is large and, at the same time, damping is high. For our example, the phase lag is $71^\circ$ at low level, where the damping ratio is 0.3 \%. The extracted modal frequency is then $0.99 \omlin{}_,{}_2$, which corresponds to a relative error of 0.9 \%. The effect on the extracted damping ratio is, however, small: The damping is extracted as 0.303 \%, instead of 0.3\%, which is a negligible error for experimental results.

\begin{figure}
	\centering
	\begin{subfigure}{0.49\textwidth}
		\centering
		\includegraphics{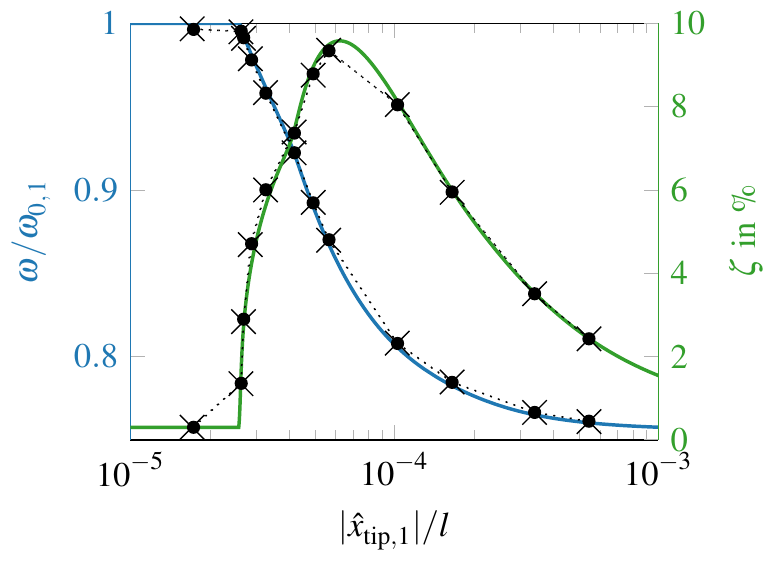}		
		\caption{}\label{fig:virtual_wvel_mode1_exc7_shaker}
	\end{subfigure}
	\begin{subfigure}{0.49\textwidth}
		\centering
		\includegraphics{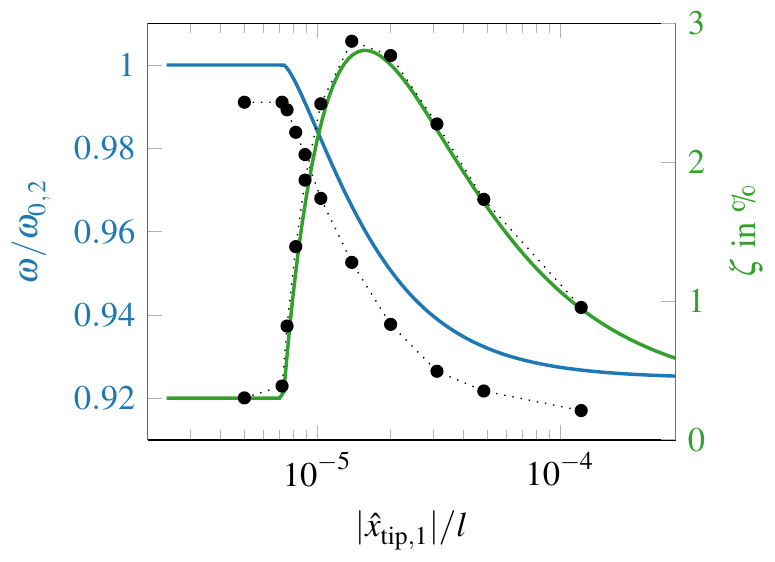}	
		\caption{}\label{fig:virtual_wvel_mode2_exc7_shaker}
	\end{subfigure}
	\begin{subfigure}{0.49\textwidth}
		\centering
		\includegraphics{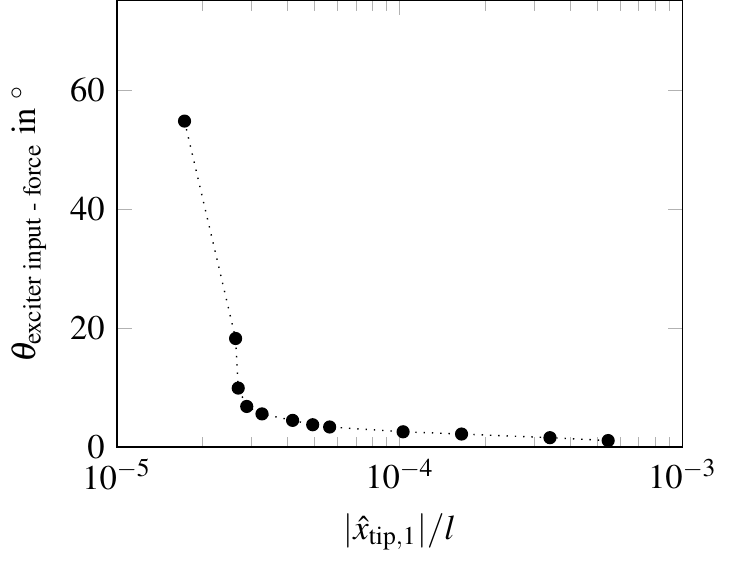}
		\caption{}\label{fig:virtual_wvel_mode1_exc7_shaker_phase}
	\end{subfigure}
	\begin{subfigure}{0.49\textwidth}
		\centering
		\includegraphics{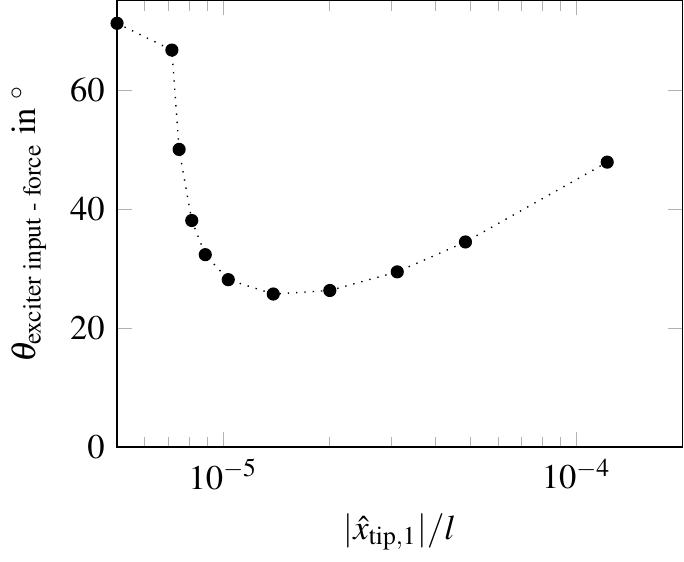}
		\caption{}\label{fig:virtual_wvel_mode2_exc7_shaker_phase}
	\end{subfigure}
	\begin{subfigure}{0.49\textwidth}
		\centering
		\includegraphics{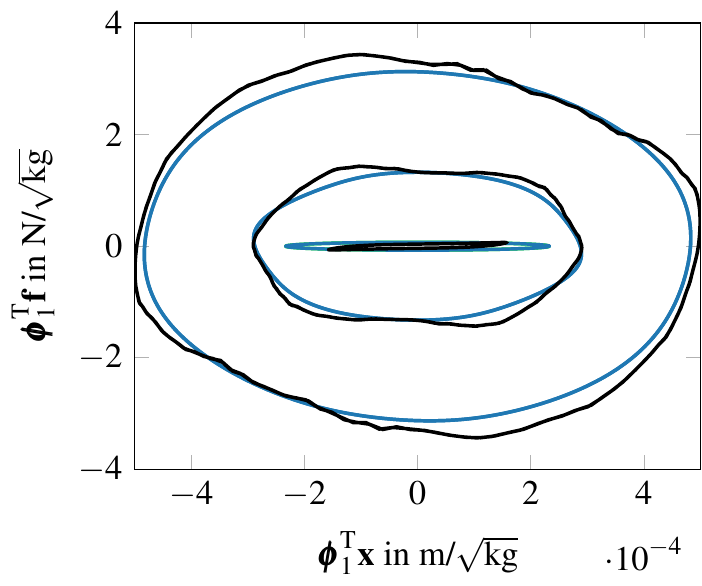}
		\caption{}\label{fig:virtual_wvel_mode1_exc7_shaker_loops}
	\end{subfigure}
	\begin{subfigure}{0.49\textwidth}
		\centering
		\includegraphics{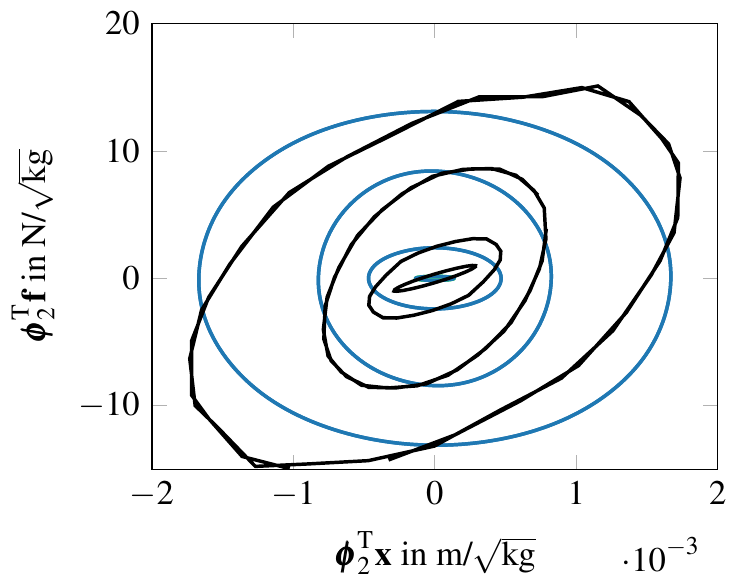}
		\caption{}\label{fig:virtual_wvel_mode2_exc7_shaker_loops}
	\end{subfigure}
	\caption{Modal properties according to the \EPMC (solid lines) and the results of the simulated experiment with multi-velocity feedback for excitation at node seven (a) for the first mode and (b) for the second mode. Dots indicate results with exciter model, crosses indicate results with exciter model and measurement noise. (c), (d) Phase lag between exciter input and excitation force for the first and second mode, respectively. (e), (f) Modal force over modal deflection for some representative levels for the first and second mode, respectively (neglecting measurement noise). Blue is the modal motion according to the \EPMC, black is the motion in the simulated experiment.}
\end{figure}

\subsection{Robustness against measurement noise}\label{sec:virtual_noise}

Finally, measurement noise is simulated in the experiment for the first mode. To this end, band-limited white noise is added to the response, thus noisy velocity signals are fed back. Additionally, noise is added to the recorded force signal to  investigate the effect of noise on the signal analysis. 
The correlation time of the band-limited white noise is set to $0.5 \cdot 10^{-3}$ s. The height of the power spectral densities are $10^{-9}$ (m/s)$^2$/Hz and $10^{-7}$ N$^2$/Hz for the response noise and force, respectively. This leads to signal to noise ratios for the force signal of 1.0 dB and 30.7 dB (lowest and highest vibration level, respectively). For the velocity signal at the excitation point, the resulting signal to noise ratios are 11.7 dB and 36.8 dB (lowest and highest vibration level, respectively).	
Weighted velocity feedback is robust against noise: The extracted modal properties are of similar accuracy than the properties without added noise (see crosses in \fref{virtual_wvel_mode1_exc7_shaker}).

\section{Experimental comparison of multi-velocity feedback and phase resonance testing}\label{sec:rubber}

In this chapter, experimental results obtained with the setup RubBeR \cite{Scheel2020b} are presented, a friction-damped cantilevered beam. The beam has six pockets with inserts. Two fixed metal plates are inserted to the second pocket from the clamping (see \fref{rubber}). The contact normal load is applied by means of an air pillow between the two metal plates. When the beam vibrates horizontally, the contact interactions between plate and beam cause a dry friction nonlinearity. For a detailed description of RubBeR, the interested reader is referred to \cite{Scheel2020b}. In this work, the pillow was filled with 0.38 bar ($38 \cdot 10^3$ Pa). With this setup, the beam's dynamics can be studied ranging from full stick to nearly full slip.

\begin{figure}
	\centering
	\begin{subfigure}[b]{0.6\textwidth}
		\centering
		\def\svgwidth{\textwidth}
		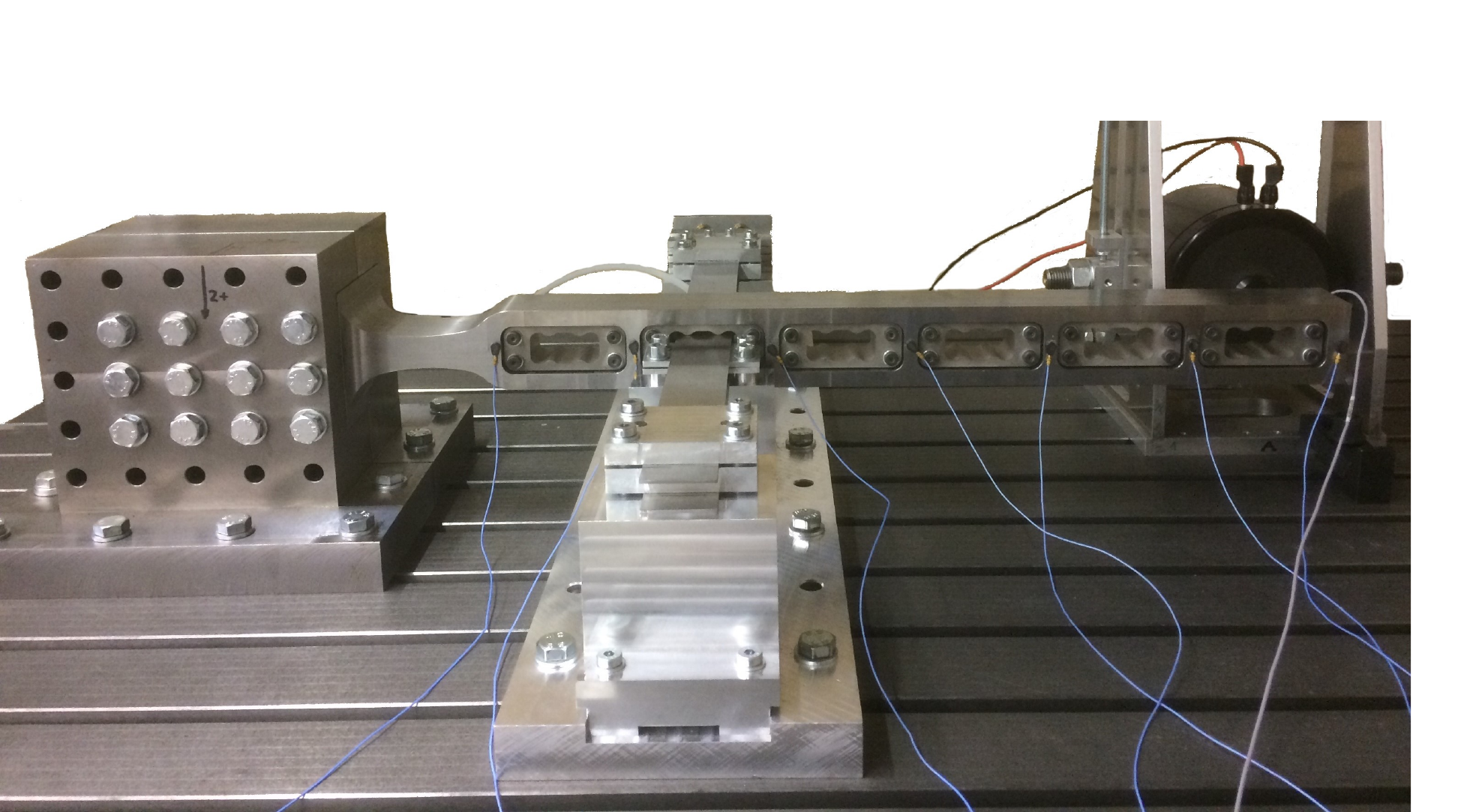
		\caption{}\label{fig:friction_nonlinearity}
	\end{subfigure}
	\begin{subfigure}[b]{0.38\textwidth}
		\centering
		\includegraphics[width=0.9\textwidth]{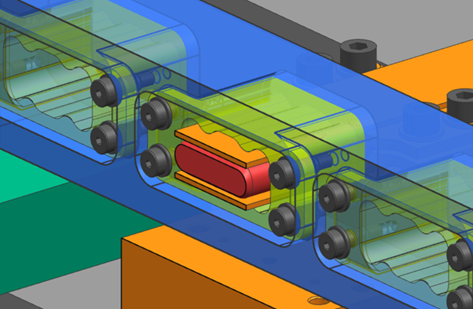}	
		\caption{}\label{fig:airpillow}
	\end{subfigure}
\caption{(a) Picture of the RubBeR setup with instrumentation. (b) CAD model detail of the friction interface with steel plates (orange), air pillow (red), inserts (green) and beam (blue) \cite{Scheel2020b}.} \label{fig:rubber}
\end{figure}

Seven accelerometers (PCB 352C22) were super-glued to the beam at half height (highlighted with white circles in \fref{rubber}). On the opposite side of the beam, a Br\"{u}el \& Kj\ae r 8230 force transducer was screwed to the beam, close to the free end. The excitation was applied via a steel stinger (3 mm diameter, 190 mm long) and a Br\"{u}el \& Kj\ae r vibration exciter type 4809. The exciter was driven by a  Br\"{u}el \& Kj\ae r power amplifier type 2719 in current mode. 

First, the beam was excited with a low-level periodic random signal from 10 Hz to 1250 Hz, using m+p international Analyzer with the hardware module VibRunner. The first three linear horizontal bending modes were identified with linear modal analysis, with the frequencies and modal damping ratios given in \tref{linear_test}. 

\begin{table} 
	\centering
	\begin{tabular}{l | l l}
		\toprule
		& $\omlin$ in Hz & $\Dlin$ in \% \\
		\midrule[0.8pt] 
		first mode & 107.2 & 0.15 \\
		second mode & 535.4 & 0.13 \\
		third mode & 999.5 & 0.44 \\
		\bottomrule
	\end{tabular}
	\caption{Linear modal frequencies and damping ratios of the first three horizontal bending modes with contact at the joint, identified with periodic random excitation at low level.} \label{tab:linear_test}
\end{table}

\subsection{Nonlinear modal testing}

In this section, the results of nonlinear modal testing using phase resonance testing and multi-velocity feedback are compared. The schemes for PLL control and velocity feedback (\cf \fref{scheme_exper_flow}) were implemented on a dSPACE MicroLabBox. Details on the PLL control parameters are given in \aref{pll}. For velocity feedback, the measured acceleration signals were integrated in real time to obtain velocity information. To compute the moving rms value, the time constant of the first order low-pass filter was set to $5 \cdot 10^{-3}$ s.
The sampling frequency in all experiments was 10,000 Hz, which is high enough to sample one period of the first mode with at least 90 points.

The modal frequency and damping ratio of the first horizontal bending mode is extracted for vibration levels ranging from 0.1 N to 32.8 N fundamental harmonic amplitude (see \fref{exp_freqdamp}). The properties are plotted over the fundamental harmonic amplitude of the beam's tip deflection, normalized with the beam's free length $l= 0.71$ m. The average of six measurements (three with increasing, three with decreasing excitation level) are indicated with a blue solid line and a green dashed line for phase resonance testing and velocity feedback, respectively. The filled areas indicate the spread of the six measurements.
The modal frequency drops by 37 \% in the measured vibration range. The modal frequency extracted with velocity feedback is lower than the extracted frequency with phase resonance testing for most vibration levels. The extracted frequencies overlap only for very low and large vibration levels. 
It is observed that the spread of phase resonance measurement data is larger than for velocity feedback.
The extracted modal damping ratio increases significantly from 0.2 \% to a maximum damping ratio of 14.7 \% for phase resonance testing. In this same amplitude range, the damping ratio extracted with velocity feedback is higher with a maximum of 15.7 \%. For large vibration levels, the damping ratio of both approaches agree well and decrease to about 3\%.
The nonlinear deflection shapes obtained with the two approaches agree similarly well, but are not visualized in this work for brevity. The interested reader is referred to \cite{Scheel2020b}, where a similar configuration was studied.

\begin{figure}
	\centering
	\begin{subfigure}{0.49\textwidth}
		\centering
		\includegraphics{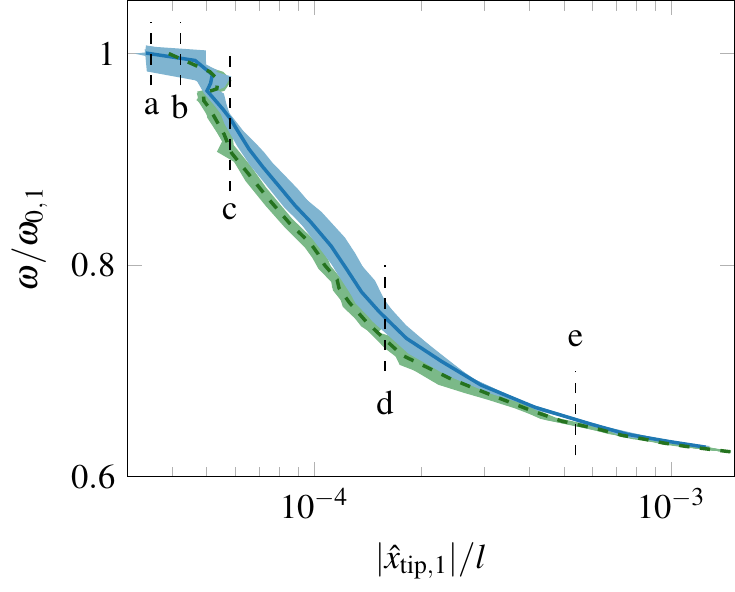}
		\caption{}
	\end{subfigure}
	\begin{subfigure}{0.49\textwidth}
		\centering
		\includegraphics{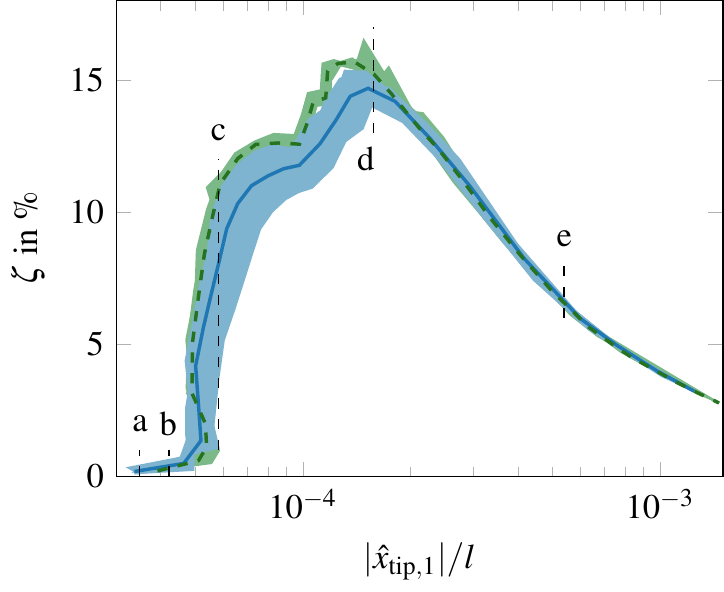}
		\caption{}
	\end{subfigure}
	\caption{(a) Modal frequency and (b) modal damping ratio of the first horizontal bending mode of RubBeR. The blue solid and green dashed line indicate the average value of phase resonance testing and velocity feedback measurements, respectively. The filled areas indicate the spread of the measurement data.}\label{fig:exp_freqdamp}
\end{figure}

\begin{figure}
	\centering
	\includegraphics{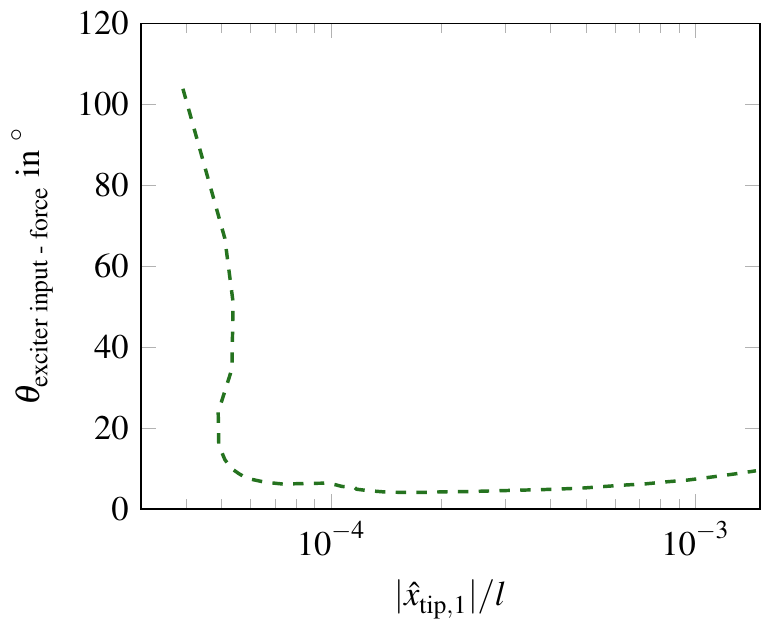}
	\caption{Phase lag between the velocity feedback excitation signal and measured excitation force.}\label{fig:exp_phaselag}
\end{figure}

The phase lag between the measured excitation force and the excitation signal, \ie the output of the MicroLabBox, is shown in \fref{exp_phaselag}. As in the simulated experiments, the phase lag is very high for low vibration levels and then drops to less than ${10}^\circ$.
It is possible that the phase lag causes the lower extracted modal frequency of velocity compared to phase resonance testing. Further, the damping ratios extracted with velocity feedback are consistently higher. Applying the analysis in \aref{phaselag}, however, the errors due to phase lag are expected to be of similar order than the spread of the measurement data. Another likely cause of the observed deviations are the gyroscopic forces introduced with multi-velocity feedback.

\subsection{Validation of modal properties with frequency response curves}

To validate the nonlinear modal properties, frequency responses to harmonic forcing are synthesized based on the nonlinear, single modal oscillator,
\e{
	[- \Omega^2+2 \ii \Omega \ommod(\modamp) \Dmod(\modamp) + \ommod(\modamp)^2] \modamp \ee^{\ii\phasemod} = \shpmodnorm\herm_1 (\modamp) \ForceVec_{1}.
}{eq:nmsdof}
$\ForceVec_{1}$ and $\Omega$ are the fundamental Fourier coefficient and angular frequency of the excitation force. $\modamp \ee^{\ii\phasemod}$ is the complex modal amplitude.
This equation is solved explicitly for $\Omega^2$ at amplitude levels covered by the backbone measurements \cite{Schwarz2019}.
To achieve a fine resolution of the frequency curve, the modal properties are first interpolated. Due to the occurrence of turning points in the modal frequency and damping ratio with respect to modal amplitude, the properties are interpolated over the arc length. To this end, the arc length $s$ is computed as Euclidean distance in the $a$-$\ommod$-$\Dmod$-space. Then, the modal properties $a(s)$, $\ommod(s)$, $\Dmod(s)$, and $\shpmodnorm (s)$ are interpolated using piecewise cubic Hermite polynomials. 
In \fref{exp_synthesis}, synthesized curves for four levels are shown for the averaged modal properties of phase resonance testing (blue solid lines) and velocity feedback (green dashed lines). Further, the filled areas indicate the spread of synthesized curves, based on the spread of the six measurements. Note that the green area is displayed on top of the blue area. Around the resonance, the spread of the velocity feedback method is smaller than the spread of phase resonance testing. This is consistent to the spread observed for the modal properties. The fundamental harmonic excitation amplitudes of the four levels are 1 N, 10 N, 14.8 N, and 26.7 N.

The synthesized frequency response curves are compared with measured reference curves.
To this end, the PLL controller is utilized to step through the phase lag around resonance (see \aref{pll} for more details). The fundamental harmonic amplitude of the excitation force was controlled to the above given values with an additional outer control loop using a PI controller with gains $K_\mathrm{i} = 0.5$ V/(Ns) and $K_\mathrm{p} = 0.5 $ V/N.
For each level, three frequency curves were measured and the average is shown \fref{exp_synthesis} (black dots). 

For all vibration levels, the reference curves agree well with the synthesized curves of the modal oscillator based on phase resonance testing. Therefore, it is concluded that these modal properties are extracted with high accuracy. The synthesized curves of multi-velocity feedback are shifted to lower frequencies (in accordance to \fref{exp_freqdamp}), especially for the two medium forcing levels.
Implementing multi-velocity feedback with the proposed single-point excitation (which leads to gyroscopic forces) and the used exciter (which introduces a considerable phase shift), it is concluded that phase resonance testing leads to results of higher accuracy than multi-velocity feedback.

\begin{figure}
	\centering
	\includegraphics{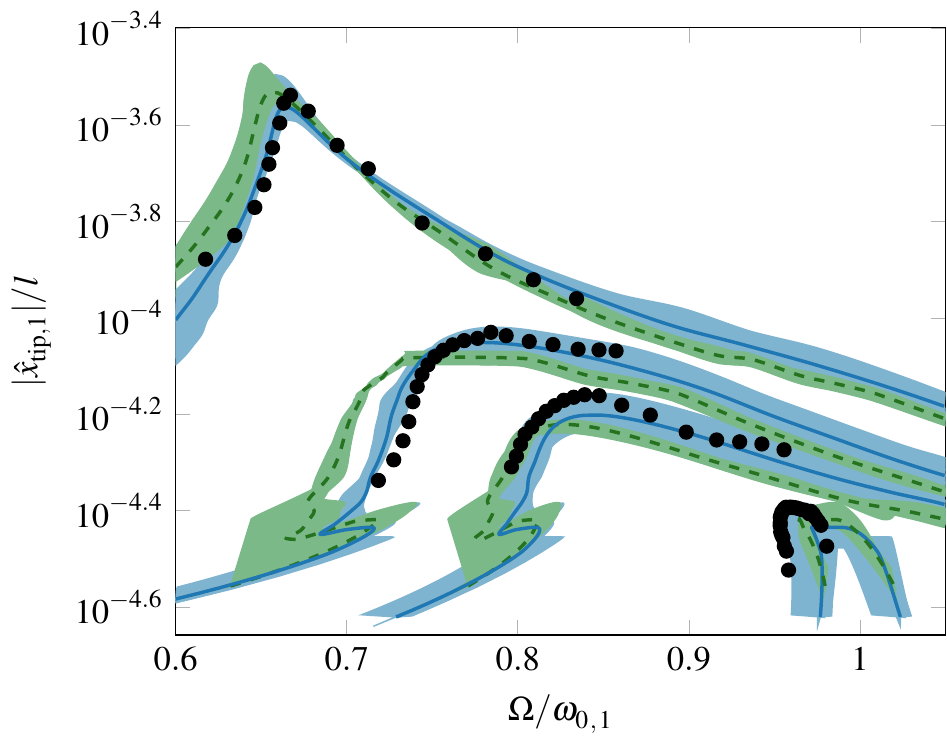}
	\caption{Synthesized (lines) and measured frequency response curves (black dots). Blue solid lines and green dashed lines indicate curves for the averaged modal oscillator obtained with phase resonance testing and velocity feedback, respectively. The blue-filled and green-filled areas indicate the spread of synthesized curves, based on the spread of the six phase resonance and velocity feedback measurements, respectively.} \label{fig:exp_synthesis}
\end{figure}


\section{Conclusions}

In this work, it is shown how velocity feedback can be implemented in an experiment with single-point excitation and utilized for nonlinear modal testing based on the \EPMC. Velocity feedback is inherently a multi-harmonic excitation (as is the negative damping term of \EPMC).
No controller is required for velocity feedback. Standard sensors, \eg accelerometers can be used in combination with real-time integration to obtain velocity information, avoiding the need of costly velocity measurements. Unstable motion can be stabilized by scaling the excitation with the reciprocal of the vibration amplitude. If the amplitude measure is determined using a low-pass filter, it is crucial that the time constant is sufficiently small.

Two approaches are derived and compared: single-velocity feedback, where only the drive-point velocity is fed back and multi-velocity feedback, where a mass-weighted sum of velocities are fed back.
Single-velocity feedback is sensitive to the choice of exciter location, \ie the responding mode depends on the exciter location (even for linear systems). Further, this approach suffers from unintended modal coupling of all modes.
With multi-velocity feedback, every mode of interest can be isolated with a given exciter location. This holds as long as the exciter is not located at a vibration node and sufficiently many linear modes are included in the estimation of the mass matrix.
Weighted velocity feedback, applied at one exciter location, causes gyroscopic forces, which distorts the modal motion.

In the present work, the accuracy of the modal properties extracted with multi-velocity feedback and phase resonance testing is compared. The results suggest that phase resonance testing, controlling the phase lag of only one harmonic, is more robust against excitation imperfections and leads to more accurate results than the proposed velocity feedback scheme. The findings of this study are based on the usage of a standard electromagnetic exciter, which introduces a non-negligible phase lag. This impedes the excitation with velocity feedback, but does not affect phase resonance testing.
The imperfections introduced due to the exciter and single-point excitation can cause large errors, which outweigh potential benefits caused by the multi-harmonic nature of velocity feedback.
The more accurate results of phase resonance testing using a PLL controller come at the cost of tuning of the control parameters.
This work focuses on friction-damped systems only. The harmonic content of the excitation might be less significant for such systems than for systems with \eg a cubic stiffness.

In future studies, the comparison could be extended by employing a phase-preserving exciter system. Multi-velocity feedback could be improved using multi-point excitation, which counteracts gyroscopic effects and reduces modal coupling.
Another objective for future research is to reduce mode coupling, caused by velocity feedback with single-point excitation. One option is to simultaneously apply positive damping to the neighboring modes through velocity feedback of opposite sign.


\section*{Acknowledgment}
 
This work was funded by the Deutsche Forschungsgemeinschaft (DFG, German Research Foundation) [Project 402813361]. The author would like to thank Malte Krack for fruitful discussions on velocity feedback and its interpretation.

\appendix
\setcounter{figure}{0}
\setcounter{table}{0}

\section{PLL implementation}\label{append:pll}

A sketch of the PLL controller and the parameters used in this work are given in \fref{pll_scheme} and \tref{pll_parameters}, respectively. As in \cite{Scheel2020b}, synchronous detection was implemented for the phase detector using a first-order low-pass filter with the time constant $2/\pi \; \mathrm{s}$.

\begin{figure}
	\centering
	\includegraphics{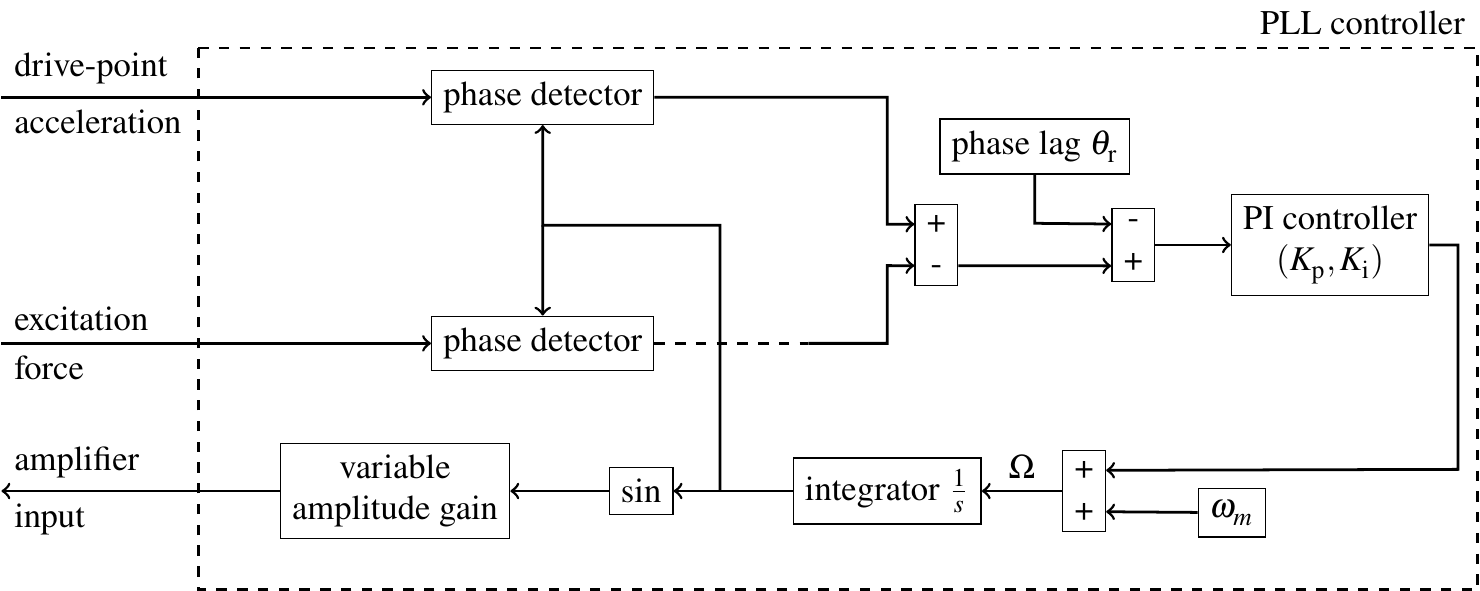}
	\caption{Scheme of the PLL controller with synchronous detection.}
	\label{fig:pll_scheme}
\end{figure}

\begin{table}
	\centering
	\begin{tabular}{c |c c c c c}
		\toprule
		measurement & \makecell{excitation level \\ (fundamental harmonic)} & $K_\mathrm{p}$ in 1/s & $K_\mathrm{i}$ in 1/s$^2$ & $\phaselag_\mathrm{r}$  & $\omega_m$ in rad/s\\
		\midrule[0.8pt] 
		backbone & 0.1 N ... 32.8 N & 5 & 15 & $90^\circ$ & 673.3\\
		\midrule[0.8pt]
		\multirow{ 4}{*}{\makecell{frequency response \\ curves}} & 1 N &  15 & 5 & $40^\circ ... 160^\circ$ & $210 \pi $\\
		 & 10 N & 5 & 15 & $55^\circ ... 145^\circ$ & $190 \pi$\\
		 & 14.8 N  & 5 & 15 & $65^\circ ... 160^\circ$ & $190 \pi $\\
		 & 26.7 N & 5 & 15 & $55^\circ ... 130^\circ$ & $178 \pi$\\
		\bottomrule		
	\end{tabular}
	\caption{Parameters of the PLL controller for measurements with RubBeR.}\label{tab:pll_parameters}
\end{table}

\section{Stability analysis using perturbation calculus}\label{append:perturbation}

The following development is inspired by the perturbation calculus analysis for lightly damped structures in \cite{GeradinRixen2015}.
Let us consider the linearized system for small vibrations around the considered equilibrium point
\e{
	\mass \acceleration + \damping \velocity + \stiffness \displacement = \forceVec = \scale \mathbf{W} \velocity }{linear_eqm}
with symmetric stiffness matrix $\stiffness$ and modal damping matrix $\damping$. The system is excited with a general form of velocity feedback.
Small modal damping is assumed, $\damping = \epsilon \damping_s$ with $\epsilon$ a small parameter. Since the focus of this work lies on vibrations close to resonance, the excitation is also assumed small, $\scale = \epsilon \scale_s$.

The solution is assumed in the form $\displacement = \mathbf{z} \ee^{\lambda \timevar}$ with $\mathbf{z} = \mathbf{z}_0 + \epsilon \mathbf{z}_1 + \mathcal{O}(\epsilon^2)$ and $\lambda = \lambda_0 + \epsilon \lambda_1 + \mathcal{O}(\epsilon^2)$.
The zero-order analysis ($\epsilon^0$) yields the eigenvalue problem of the conservative system
\e{
	\left( \lambda_0^2 \mass + \stiffness \right) \mathbf{z}_0 = \mathbf{0}
}{}
with $\lambda_0 = \pm \ii \omlin{}_,{}_i$ and $\mathbf{z}_0 = \shpmodnormlin_i$.
The first-order analysis ($\epsilon^1$) and left multiplication with $\mathbf{z}_0\tra$ yields
\e{
	\pm \ii \omlin{}_,{}_i \left( 2 \lambda_1 + 2 \Dlin{}_,{}_i \omlin{}_,{}_i - \scale_s \shpmodnormlin_i\tra \mathbf{W} \shpmodnormlin_i  \right) = 0
}{}
and thus 
\e{
	\lambda_1 = - \left( \Dlin{}_,{}_i \omlin{}_,{}_i - \frac{1}{2} \scale_s \shpmodnormlin_i\tra \mathbf{W} \shpmodnormlin_i  \right).
}{}
$\Dlin{}_,{}_i$ is the modal damping ratio of the $i$-th mode.
The mode is negatively damped, or in other words excited, when the cumulative damping is negative. That is, $\lambda_1 > 0$ holds.

In case of single-velocity feedback, $\mathbf{W}_\text{sfb} = \unitvec\indexexc \unitvec\indexexc\tra$ and thus $\shpmodnormlin_i\tra \mathbf{W}_\text{sfb} \shpmodnormlin_i = \shpmodnormscalarlin_{i,k}^2$.
The $i$-th mode is negatively damped when $\scale_s > 2 \Dlin{}_,{}_i \omlin{}_,{}_i / \shpmodnormscalarlin_{i,k}^2$.

In case of multi-velocity feedback, $\mathbf{W}_\text{mfb} = \unitvec\indexexc \shpmodnormlin\tra\indexmode{}_,{}\exper \mass\exper \locationexp$. Then, 
\e{
\shpmodnormlin_i\tra \mathbf{W}_\text{mfb} \shpmodnormlin_i
	\begin{cases}
		= \shpmodnormscalarlin_{m,k} \quad \text{for the mode of interest (part of the } \numberdof_\mathrm{m} \text{ modes used for the estimation of } \mass\exper). \\
		= 0 \quad \quad \text{ for the other identified modes.} \\
		\neq 0 \quad \quad \text{ for a mode that is \emph{not} identified.} \\
	\end{cases}
}{}
In the first and third case, a mode is negatively damped when $\scale_s > 2 \Dlin{}_,{}_i \omlin{}_,{}_i / (\shpmodnormlin_i\tra \mathbf{W}_\text{mfb} \shpmodnormlin_i)$.
Assuming that only modes with significantly higher frequency with similar or higher damping are not included in the set of $\numberdof\indexmode$ modes, the first mode to meet that criterion is the mode of interest and thus $\scale_s > 2 \Dlin{}_,{}\indexmode \omlin{}_,{}\indexmode / \shpmodnormscalarlin_{m,k}$.

\section{Model of the electromagnetic exciter}\label{append:shaker_model}

To explain the phase lag caused by the electromagnetic exciter, the exciter is modeled with an electric circuit and a mechanical single \DOF\ oscillator as shown in \fref{scheme_exciter} \cite{McConnellVaroto2008}. The motion of the exciter's armature $x_{\rm{a}}$ is defined by the equation of motion
\e{
	m_{\rm{a}} \ddot{x}_{\rm{a}} + d_{\rm{a}} \dot{x}_{\rm{a}} + k_{\rm{a}} x_{\rm{a}} = G i - \forcescalar_\text{stinger}.
}{eqm_armature}
$i$ is the current in the electric circuit, and $\forcescalar_\text{stinger}$ is the force transmitted between exciter and structure. 
The dynamics of the structure under test is reduced to one linear mode (frequency $\omlin$, damping ratio $\Dlin$, mode shape $\shpmodnormlin$). Let us assume that the structure under test is attached to the exciter through a rigid stinger. Thus, the armature motion is equal to the structure's motion at the excitation location. The entry of the mode shape associated to that point is $\shpmodnormscalarlin\indexexc$.
With this, the structure's dynamics is defined by
\e{
	\ddot{x}_{\rm{a}} + 2 \Dlin \omlin \dot{x}_{\rm{a}} + \omlin^2 x_{\rm{a}} = |\shpmodnormscalarlin\indexexc|^2 \forcescalar_\text{stinger}
}{eqm_struct}

\begin{figure}
	\centering
	\def\svgwidth{0.7\textwidth}
	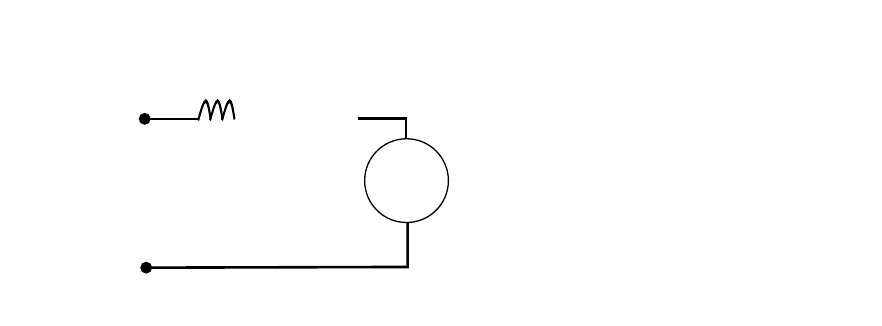
	\caption{Model of the exciter}\label{fig:scheme_exciter}
\end{figure}

The excitation signals applied in this work are dominated by the fundamental harmonic. The following analysis is therefore simplified by considering mono-harmonic signals only.

Transforming \eref{eqm_armature} and \eref{eqm_struct} to frequency domain yields
\ea{
	(-m_{\rm{a}} \Omega^2 + \mathrm{i}\Omega d_{\rm{a}}  + k_{\rm{a}}) \hat{x}_{\rm{a}} &= G \hat{i} - \hat{\forcescalar}_\text{stinger}\\
	(-\Omega^2 + \mathrm{i}\Omega 2 \Dlin \omlin + \omlin^2) \hat{x}_{\rm{a}} &= |\shpmodnormscalarlin\indexexc|^2 \hat{\forcescalar}_\text{stinger}.
}{}

If the power amplifier is operated in current mode, the current $i$ is assumed to be proportional to the excitation voltage $u$ up to a break frequency \cite{McConnellVaroto2008}. The dynamic transfer behavior from exciter input to excitation force is then defined by
\e{
	\dfrac{\hat{\forcescalar}_\text{stinger}}{G \hat{i}}= \dfrac{(-\Omega^2 + \mathrm{i}\Omega 2 \Dlin \omlin + \omlin^2) }{ (-\Omega^2 + \mathrm{i}\Omega 2 \Dlin \omlin + \omlin^2) + |\shpmodnormscalarlin\indexexc|^2 (-m_{\rm{a}} \Omega^2 + \mathrm{i}\Omega d_{\rm{a}}  + k_{\rm{a}})}.
}{}
The transfer behavior depends on the attached structure under test. If the excitation frequency $\Omega$ is equal or close to the structure's resonance (for example in case of phase resonance testing), the transfer function simplifies to

\e{
	\dfrac{\hat{\forcescalar}_\text{stinger}}{G \hat{i}}(\Omega \approx \omlin) \approx \dfrac{ 1}{ 1 + \dfrac{|\shpmodnormscalarlin\indexexc|^2 m_{\rm{a}}}{\mathrm{i} 2 \Dlin} \left(-1 + \mathrm{i}2 \zeta_{\rm{a}} \dfrac{\omega_{\rm{a}}}{\omlin}  + \left(\dfrac{\omega_{\rm{a}}}{\omlin} \right)^2 \right)}
}{}
with the armature's modal frequency $\omega_{\rm{a}} = \sqrt{k_{\rm{a}} / m_{\rm{a}}}$ and damping ratio $ \zeta_{\rm{a}} = d_{\rm{a}} /( 2 m_{\rm{a}} \omega_{\rm{a}}$).

If the armature's and structure's modal frequency coincide, the transfer function is real. For nonlinear systems, whose modal frequency change with the vibration level, these two frequencies do not coincide for all vibration levels. Then, the transfer function is complex, especially for lightly damped structures. Thus, the exciter-structure interaction causes a phase lag between exciter input and excitation force.

The exciter's parameters (see \tref{exciter_parameters}) are identified following \cite{DellaFlora2008} for a Br\"{u}el and Kj\ae r vibration exciter type 4809, which is the device used in the experiments.

\begin{table}
	\centering
	\begin{tabular}{l|l}
		\toprule
		Parameter & Value\\
		\midrule[0.8pt] 	
		$m_{\rm{a}}$ & 0.057 kg\\
		$k_{\rm{a}}$ & 9932 N/m\\
		$d_{\rm{a}}$ & 21.51 Ns/m\\
		$G$ & $6.78 $ N/A\\
		$L$ & $140 \cdot 10^{-6}$ H\\
		$R$ & $2 \text{ }\Omega$\\
		$k_{\text{stinger}}$ & $4.4\cdot 10^{7} $ N/m\\
		\bottomrule
	\end{tabular}
	\caption{Parameters of the exciter, used in the simulated experiment.}\label{tab:exciter_parameters}
\end{table}

\section{Effect of a phase lag on the extracted modal properties of a linear system}\label{append:phaselag}

Let us consider the externally forced, linear system of \eref{linear_eqm} with modal damping. The modal properties of one mode are $\omlin$, $\Dlin$ and $\shpmodnormlin$. A single-point, harmonic force is applied at the $k$-the \DOF , $\forceVec = \unitvec\indexexc \forcescalar$, with frequency $\Omega$ and amplitude $\hat{\forcescalar}$.

If a mode is perfectly isolated, the system vibrates in resonance with $\Omega = \omlin$. Then, all modal properties are extracted accurately with the procedure described in \sref{enma}. If the excitation deviates from the ideal resonance case due to a phase lag, the system vibrates with an off-resonant frequency, $\Omega = \Omega_{\text{err}} \neq \omlin$.
	
Let us consider the projection of the off-resonant vibration onto the mode. For a given phase lag $\theta$ between drive-point velocity and excitation force, the excitation frequency (and thus the extracted modal frequency) is 
\e{
	\ommod_{\text{err}} = \Omega_{\text{err}} = \omlin \left( \sqrt{\Dlin^2 \tan^2(\theta) +1} - \Dlin \tan(\theta) \right).
}{}
The magnitude of the modal amplitude is then given as
\e{
	\modamp^2 = \dfrac{|\hat{\forcescalar}|^2 \shpmodnormscalarlin\indexexc^2}{(\omlin^2-\Omega_{\text{err}}^2)^2 + (2 \Dlin \omlin \Omega_{\text{err}})^2}.
}{}
With the active power
\e{
	P = \dfrac{|\hat{\forcescalar}|^2 \shpmodnormscalarlin\indexexc^2 \Dlin \omlin \Omega_{\text{err}}^2}{(\omlin^2-\Omega_{\text{err}}^2)^2 + (2 \Dlin \omlin \Omega_{\text{err}})^2},
}{}
the modal damping ratio is extracted as
\e{
	\Dmod_{\text{err}} = \dfrac{\Dlin}{\sqrt{\Dlin^2 \tan^2(\theta) +1} - \Dlin \tan(\theta)}.
}{}

\bibliography{VelocityFeedback_paper_ref}

\end{document}